\documentclass[apj,numberedappendix,appendixfloats,iop]{emulateapj}

\usepackage%
[
pagebackref, % or backref
colorlinks=true,
linkcolor=red,filecolor=red,
citecolor=blue,
urlcolor=cyan,%------------- Doc Info ---------------------------------
pdftitle={Deuterium Burning and Core-Nucleated Accretion},
pdfauthor={Bodenheimer et al.},
pdfsubject={Deuterium Burning and Core-Nucleated Accretion},
pdfcreator={KAZEditor},
pdfproducer={Kazzimiei},
pdfkeywords={planet formation; giant planets; brown dwarfs; extrasolar planets},
bookmarks=true,
breaklinks=true,
hypertexnames=false,
pdfstartview=FitV,
hyperfootnotes=true,
bookmarksopen=false,
pdfpagemode=None]{hyperref}
\usepackage{breakurl}
\usepackage{amsmath}

\long\def\symbolfootnote[#1]#2{\begingroup%
\def\thefootnote{\fnsymbol{footnote}}\footnote[#1]{#2}\endgroup} 

\slugcomment{Accepted for publication in The Astrophysical Journal}

\shorttitle{Deuterium Burning and Core-Nucleated Accretion}
\shortauthors{Bodenheimer, D'Angelo, Lissauer, Fortney, and Saumon}

\begin{document}


\title{Deuterium Burning in Massive Giant Planets and Low-Mass Brown Dwarfs
 formed by Core-Nucleated Accretion}


%\author{Peter Bodenheimer}
%\affil{UCO/Lick Observatory, Department of Astronomy and Astrophysics, University of California,
%    Santa Cruz, CA 95064}
%\email{peter@ucolick.org}
%
%\author{Gennaro D'Angelo\altaffilmark{1,2} and Jack J. Lissauer}
%\affil{Space Science and Astrobiology Division, NASA-Ames Research Center, 
%Moffett Field, CA 94035}
%\email{gennaro.dangelo@nasa.gov}
%\email{Jack.J.Lissauer@nasa.gov}
%
%\author{Jonathan  J. Fortney}
%\affil{Department of Astronomy and Astrophysics, University of California,
%    Santa Cruz, CA 95064}
%\email{jfortney@ucolick.org}
%
%\and
%
%\author{Didier Saumon}
%\affil{Los Alamos National Laboratory, P. O. Box 1663, Los Alamos, NM 87545}
%\email{dsaumon@lanl.gov}
%
%\altaffiltext{1}{SETI Institute, 189 Bernardo Avenue, Mountain View, CA 94043}
%\altaffiltext{2}{Visiting Research Scientist, Los Alamos National Laboratory, Los Alamos, NM 87545}

\author{Peter Bodenheimer\altaffilmark{1}, Gennaro D'Angelo\altaffilmark{2,4,5}, 
            Jack J. Lissauer\altaffilmark{2}, Jonathan J. Fortney\altaffilmark{1},
             and Didier Saumon\altaffilmark{3}}
\altaffiltext{1}{UCO/Lick Observatory, Department of Astronomy and Astrophysics, 
                 University of California, Santa Cruz, CA 95064
                 (E-mail: \href{mailto:peter@ucolick.org}{peter@ucolick.org},
                  \href{mailto:jfortney@ucolick.org}{jfortney@ucolick.org})}
\altaffiltext{2}{Space Science and Astrobiology Division, NASA Ames Research Center, 
                 Moffett Field, CA 94035
                 (E-mail: \href{mailto:gennaro.dangelo@nasa.gov}{gennaro.dangelo@nasa.gov},
                  \href{mailto:Jack.J.Lissauer@nasa.gov}{Jack.J.Lissauer@nasa.gov})}
\altaffiltext{3}{Los Alamos National Laboratory, P. O. Box 1663, Los Alamos, NM 87545
                      (E-mail: \href{mailto:dsaumon@lanl.gov}{dsaumon@lanl.gov})}
\altaffiltext{4}{SETI Institute, 189 Bernardo Avenue, Mountain View, CA 94043}
\altaffiltext{5}{Visiting Research Scientist, Los Alamos National Laboratory, Los Alamos, NM 87545}

\begin{abstract}

Formation of bodies near the deuterium-burning limit is considered by detailed numerical simulations
according to the core-nucleated giant planet accretion scenario. The objects, with
heavy-element cores in the range 5--30~M$_\oplus$, are assumed to accrete
gas up to final masses of 10--15 Jupiter masses (M$_\mathrm{Jup}$).
After the formation process, which lasts 1--5 Myr and which ends with a
`cold-start', low-entropy configuration, the bodies evolve
at constant mass up to an age of several Gyr. Deuterium burning via 
proton capture is
included in the calculation, and we determined the mass, $M_{50}$, 
above which more
than 50\% of the initial deuterium is burned. This often-quoted
borderline between giant planets and brown dwarfs is found to depend
only slightly on parameters, such as core mass, stellar mass,
 formation location,  solid surface density in the protoplanetary disk, 
disk viscosity, and dust opacity.
The values for $M_{50}$ fall in the range 11.6--13.6~M$_\mathrm{Jup}$, in
agreement with previous determinations that do not take the formation
process into account. For a given opacity law during the formation process, 
objects with higher core masses form more quickly. The result is higher
entropy in the envelope at the completion of accretion, yielding lower
values of $M_{50}$.
 For masses above $M_{50}$, during the deuterium-burning
phase, objects expand and increase in luminosity by 1 to 3 orders 
of magnitude. Evolutionary tracks in the luminosity--versus--time
diagram are compared with the observed position of the companion to
Beta Pictoris.
\end{abstract}


\section{Introduction}
\label{sect:intro}

There is considerable debate over the
question of defining a precise boundary between the class of objects
known as `planets' and those known as `brown dwarfs'. It has been
suggested that the two types of objects could be distinguished by
their formation mechanism; however,  it is generally difficult to 
deduce this property from observations of specific objects. Nevertheless,
there is a well-defined minimum in the mass distribution (actually $M \sin i$),
for substellar companions to G and K main-sequence stars, in the range 20--30 
M$_\mathrm{Jup}$ \citep{lov06,sah11,sch11}, suggesting that objects near the deuterium-burning limit can `form like planets'. These authors suggest that this minimum
does correspond to a (somewhat imprecise) dividing line between formation 
mechanisms and that the upper limit to planet masses should be set at
about $M \sin i \approx 25$~M$_\mathrm{Jup}$. However, no break is seen 
near this mass in the distribution of free-floating objects observed in 
the Sigma Orionis young cluster \citep{pen12} down to 4~M$_\mathrm{Jup}$. 
 Together, the observations imply that formation mechanisms do not define a unique mass boundary between planets and brown dwarfs.

Another commonly
used  criterion to classify planets, brown dwarfs and stars is based on nuclear fusion that does or does not occur within the object. 
Brown dwarfs are defined to be those objects that at some point in their 
evolution become hot enough in their
 interiors to burn a majority of the deuterium that was
 initially present in the object; however, they never become hot
enough to burn $^1$H by the proton-plus-proton reaction in a self-sustaining
manner as true stars do. 
On the other hand, the term planet is applied only to objects that will not burn much deuterium. This criterion was used
by \citet{bur97} to separate the two types of objects, and the dividing
line was stated to be $\sim 13$~M$_\mathrm{Jup}$, where 
M$_\mathrm{Jup}= 1.898 \times 10^{30}$ g. This dividing line
depends on the helium mass fraction, the deuterium
abundance, and the metallicity, and \citet{spi11} found that for a 
reasonable range of parameters, 50\% of the initial D is burned in the
mass range 12--14~M$_\mathrm{Jup}$.
The evolutionary models used to establish the criterion have a
uniform chemical composition, a defined total mass in the vicinity
of the 13~M$_\mathrm{Jup}$ limit, constant in
time, an initial radius of about 2--3 R$_\mathrm{Jup}$ where
R$_\mathrm{Jup}=7.15 \times 10^9$~cm, and an
initial photospheric    temperature ($T_\mathrm{eff}$) of about
2500~K. The corresponding initial luminosities are $2-3 \times
10^{-3}$~L$_\odot$. A starting model of this type has become known
as a `hot-start' model, characterized by a relatively  high initial entropy 
\citep{mar07}.

The question of whether,
 either for brown dwarfs or planets, the formation mechanism actually
leads to such hot-start initial conditions is still under investigation.
For objects formed either  by  collapse of interstellar clouds
or by fragmentation in a protostellar disk by gravitational
instability,  it is plausible that the hot-start
initial condition could be reached \citep{bar02}. In the case of
gravitational instability, \citet{gal12} have found, from 
three-dimensional numerical simulations,  that the entropy
of newly-formed clumps, near the point where molecular 
dissociation sets in at the center,  is high, 
possibly consistent with  a hot 
start. However the dominant process
for giant planet formation is most likely the core-nucleated 
accretion mechanism, in which solid particles first
accumulate to form a heavy-element core, then later when
the core has attained roughly 5--10~M$_\oplus$, gas is captured
from the disk. A particular set of evolutionary calculations
based on this theory  \citep{mar07} shows that
once the planet has become fully formed, its entropy is
relatively low, with luminosities on the order of $10^{-5} - 10^{-6}$ 
L$_\odot$. The low entropy is a direct consequence of the assumption
made in these calculations that,  during the phase of rapid gas
accretion, all of the accretion energy is radiated away at the 
accretion shock at the planet's surface. Thus, the core accretion 
process can lead to a `cold start'. However the shock treatment is
approximate, and the accretion flow cannot actually be modelled
correctly with 1-D spherically symmetric calculations. Thus  
other possibilities can arise. \citet{mor12} show
that core accretion formation calculations in which none of the
energy is radiated at the shock lead to hot-start conditions very
similar to those assumed by \citet{bar03} and \citet{bur97}. Furthermore,
intermediate `warm' states are also possible outcomes \citep{spi12}.
In the core-accretion picture, also,  the chemical composition is not
uniform because of the presence of the core, which turns out
for the case of a Jupiter mass planet to fall in the range
4--20~M$_\oplus$ \citep{mov10}. 

A massive object of 25~M$_\mathrm{Jup}$
formed by core accretion \citep{bar08} has been shown to burn all of its
initial deuterium despite the presence of a heavy-element core of
100 or a few hundred  M$_\oplus$.  Cold-start models, including the core
and calculations of the formation phase,  have been investigated
to determine the D-burning mass limit \citep{mol12}.  The results
show that the limit still falls within the range 12--14 
M$_\mathrm{Jup}$.          The purpose of
the present paper is to present further formation calculations for bodies formed by core-nucleated accretion 
that end up with a total mass in the 10--15~M$_\mathrm{Jup}$ range in 
the low-entropy state, 
and to investigate the effect of various        possible
initial conditions, as well as physical parameters during the
formation stage,  on the corresponding deuterium-burning
limit.

\section{Computational Method}

The evolutionary calculations for giant planets are started at
the point where the heavy-element core has a mass of about
1~M$_\oplus$, and are carried through the entire formation process as
well as the subsequent contraction/cooling phase at constant mass,
up to a final age of several Gyr. The assumptions and computational
procedures were described in detail in previous publications
\citep{pol96,bod00,hub05,lis09,mov10}.
The early phase of the formation
process is dominated by the accretion of planetesimals onto
the core; during this phase the gaseous envelope has low mass, 
$\ll 1$~M$_\oplus$, and a low accretion rate compared to that of
the core. The latter is given by 
\begin{equation}
\frac{dM_{\rm core}}{dt} = \pi R^2_{\rm capt} \sigma \Omega_{p} F_g
\end{equation}
where $\pi R^2_{\rm capt}$ is the effective geometrical
capture cross section, $\sigma$ is the surface density of
solid particles (planetesimals) in the protoplanetary disk, $\Omega_{p}$
is the planet's orbital frequency, and $F_g$ is the
gravitational enhancement factor, which is obtained from the
calculations of \citet{gre92}. The planetesimal
radius is taken to be 50 km for the cases with a central star
of 2~M$_\odot$ and 100 km for the cases with a star of 1~M$_\odot$ (see 
Table~\ref{table:1}).  The  smaller size, or a reasonable distribution
of planetesimal sizes, tends to reduce the formation time
but has little effect on the basic results of this paper.

If no gaseous envelope is present, then $ R_{\rm capt}=R_{\rm core}$,
the radius of the heavy-element core. However, even if the
envelope mass is relatively small compared with the core mass,
the planetesimals interact with the envelope gas, are slowed down
by gas drag, and are subject to ablation and fragmentation. The
trajectories of planetesimals through the envelope are calculated
\citep{pod88}, and the effective  $R_{\rm capt}$ is determined.
The material that is deposited in the envelope is then allowed to
sink to the core, as discussed by  \citet{pol96}. Calculations by \citet{iar07}
show that this assumption is valid
at least for the organic and rock components of the planetesimals.
The ices, however, can dissolve in the envelope, so that our
`core mass' is somewhat overestimated; the quoted value actually
refers to the total excess of heavy-element material, above the
solar abundance, in the entire planet. Erosion of the core  and possible
mixing of some core material into the envelope is not considered. This
process has been shown to be unlikely for the case of 
Jupiter \citep{lis07}, but such estimates have not been extended to the
case of planets in the 10~M$_\mathrm{Jup}$ range.

The structure of the hydrogen-helium envelope is calculated according to
the differential equations of stellar structure \citep{kip90}, which assume
hydrostatic equilibrium, a spherically symmetric mass distribution, 
radiative or convective energy transport, and energy conservation.
The  energy sources are provided by planetesimal accretion, contraction
of the gaseous envelope, and cooling.  The additional energy
source provided by deuterium burning is included in the later
phases of accretion and during the constant-mass final cooling
phase, once the mass has exceeded   10~M$_\mathrm{Jup}$ and internal
temperatures exceed $\approx 10^5$~K. The full set of equations, 
supplemented by calculation of the mass accretion rates onto the core and
the envelope, and of the planetesimal trajectories, 
is solved by the Henyey method \citep{hen64}.

At the inner boundary of the
envelope the radius is set to $R_\mathrm{core}$, which is determined
from its mean density. During the earlier phases of the evolution,
when the envelope mass is less than  about 0.1~M$_\mathrm{Jup}$, the core is
assumed to be a mixture of rock and ice with a mean density of 
$3.0\,\mathrm{g\,cm}^{-3}$. During the later phases, when the pressure at the
base of the envelope increases to values above  
$\sim 10^{11}\,\mathrm{dynes\,cm}^{-2}$, 
an ANEOS equation of state with 50\% rock and  50\% ice
\citep{tho90}
for the core is used to determine its mean density, which can increase 
to $60\,\mathrm{g\,cm}^{-3}$ or higher.
In the  hydrogen-helium envelope, the equation of state is taken
to be given by  the tables of \citet{sau95}, which take into account
the partial degeneracy of the electrons as well as non-ideal effects.
The chemical composition is taken to be near-solar, with $X=0.70, 
~Y=0.283$, and $Z=0.017$, where $X,~Y,~Z$ are, respectively, the
mass fractions of hydrogen, helium, and heavy elements. The tables
of course do not include a $Z$ component, so the $Y$ component was
adjusted upwards to partially compensate.

 The Rosseland mean opacity during the formation phase combines
the low-temperature atomic/molecular calculation of \citet{ale94}
with the interstellar grain opacities of \citet{pol85}.
The opacity values of the grain component are
reduced by a factor 50 to approximately represent the reduction
caused by grain growth and settling in the protoplanet \citep{pod03,mov08}.
However, in two of the runs the grain growth and settling are calculated
in detail in the temperature range 100--1800~K as described in \citet{mov10}.
The grain size distributions
and the opacities are recalculated in every layer at every time step in that
temperature range.  These opacities are important in
regulating the rate at which the envelope can contract, and therefore
the rate at which it accretes gas. However, once the envelope is
well into the rapid gas accretion phase, at about 0.25~M$_\mathrm{Jup}$,
the gas accretion rate is limited by the physical properties of
the protoplanetary disk near the planet, and the precise values
of envelope opacity assume a less-important role. Once the planet
reaches its final mass, say 12~M$_\mathrm{Jup}$, the grains are
assumed to settle rapidly and to evaporate in the interior. For
the final contraction/cooling phase at constant mass, the molecular
opacities of \citet{fre08} are used, with solar composition, 
up to a temperature of 3500~K. At and above that temperature, with
any reasonable opacity, the interior is convective.

 At the outer surface of the envelope,  the mass addition
rate of gas, during the earlier phases of accretion, 
 is determined by the requirement that the planet radius $R_p$
match the effective accretion radius, which is given by
\citep{lis09}           
\begin{equation}
R_{\rm eff} = \frac{GM_p}{c_s^2 + \frac{GM_p}{KR_H}}~,
\label{eq:reff}
\end{equation}
where $c_s$ is the sound speed in the disk, $R_H$ is the 
Hill sphere radius, and $M_p$ is the total mass of the planet.
The constant $K \approx 0.25$ is determined by three-dimensional
numerical simulations which calculate the accretion rate of
gas from the  protoplanetary disk onto the planet \citep{lis09}.
As a result, in the
limit where $R_H$ is  small compared with the Bondi
accretion radius $GM_p/c_s^2$, $R_{\rm eff}= 0.25 R_H$.

Additional boundary conditions at the surface depend on the
evolutionary phase. During the early phases when $M_p < 0.25 
M_\mathrm{Jup}$,  the density and temperature are set to constant 
values appropriate for the protoplanetary disk, $\rho_\mathrm{neb}$ and
$T_\mathrm{neb}$, respectively.  The density $\rho_\mathrm{neb}$ is 
determined from the assumed value of $\sigma$ 
using a standard gas-to-solid ratio of 70 and 
$H_p/a_p=0.05$, where $H_p$ is the (gaussian) disk scale height and $a_p$ is the
distance of the planet from the star. However at some point during the
rapid gas accretion phase,  the mass addition rate required by
condition (\ref{eq:reff}) exceeds the rate at which matter can be
supplied by the disk. The disk-limited rates, based on three-dimensional
hydrodynamic simulations, are described in the next section. 
During that phase, the boundary conditions at the actual surface
of the planet, whose radius falls well below $R_{\rm eff}$,
  are determined through the properties of the
accretion shock at this surface, as described in detail by \citet{bod00}.
The basic assumption is that practically all of the gravitational energy
released by the infalling gas is radiated away at the shock; this energy
escapes through the infalling envelope ahead of the shock. This assumption
defines the `cold start' for planetary evolution.

 During the final phase of cooling at
constant mass, the planet becomes isolated from the disk and
the surface boundary conditions change again, to those of
a blackbody in hydrostatic equilibrium
\begin{equation}
L=4 \pi R_p^2 \sigma_B T_{\rm eff}^4 ~~~~~{\rm and}~~~~~~
\kappa P = \frac{2}{3} g~,
\end{equation}
where $\sigma_B$ is the Stefan-Boltzmann constant, $T_{\rm eff}$
is the surface temperature, $L$ is the total luminosity, 
and $\kappa$, $P$, and $g$
are, respectively, the photospheric values of Rosseland mean opacity, 
 pressure, and acceleration of gravity.
Insolation from the star is not included.

Significant deuterium burning in the mass range considered begins
near the end of the phase of rapid gas accretion. The burning occurs
via the reaction
\begin{equation}
^2\textrm{D} +~^1\textrm{H} \rightarrow~^3\textrm{He} + \gamma
\end{equation}
with an energy release $Q_\mathrm{dp} = 5.494$ MeV per reaction.
The initial deuterium abundance by mass fraction is set to $4 \times 10^{-5}$,
consistent with the value derived from the local interstellar medium
\citep{pro10}.      The reaction rate (reactions per second per gram)
is taken from the
Nuclear Astrophysics Compilation of Reaction Rates \citep{ang99}:
\begin{eqnarray}
& & R_\mathrm{dp}(\rho,T_6) = \frac{ 5.365 \times 10^{28} \rho X_{1H} 
X_{2H}}{T_{6}^{2/3}}\exp(-37.21/T_{6}^{1/3}) \nonumber \\
& &\{1+T_6 [0.0143 + T_6(3.95 \times 10^{-7} T_6 -9.05 \times 10^{-5})]\},
% RDP(RHO,T6,T63)= 5.365e+28*RHO*X1H*X2H/T63**2*exp(-37.21/T63)*\break
% (1.d0+ T6*(0.0143 + T6*(-9.05e-05 + 3.95e-07*T6)))~,
\end{eqnarray}             
where $T_6$ is the  temperature in $10^6$~K, $\rho$ is the density in cgs,
  $X_{1H}$ is the mass fraction of $^1$H, and $X_{2H}$
 is the
mass fraction of $^2$H (deuterium). This rate is then multiplied
by the screening factor, which takes into account ion-ion and ion-electron
screening in partially degenerate  dense plasmas \citep{pot12}. 
The energy generation $\epsilon$, per gram
per second, is then obtained, zone by zone,  from the rate multiplied by 
$Q_\mathrm{dp}$ in the appropriate units. To get the change in the deuterium
abundance during one time step, it is assumed that the planet interior is
fully convective and therefore fully mixed. This assumption is valid for
the planets considered during the phase of contraction and cooling, even
if no deuterium is burned. 
The convective velocities of order 10-100~$\mathrm{cm\,s}^{-1}$, 
calculated according to the mixing-length approximation, give a mixing time 
scale far shorter than the D-burning time scale. The reaction
rate multiplied by zone mass is integrated over the entire envelope and
used to calculate the abundance change.

Given the central stellar mass $M_\ast$, the solid surface density  $\sigma$,
and the distance of the planet from the star $a_{p}$, the isolation
mass for the solid material is
\begin{equation}
M_\mathrm{iso} = \frac{8}{\sqrt{3}}(\pi C)^{3/2} M_\ast^{-1/2} \sigma^{3/2} a_{p}^3
\label{eq:iso}
\end{equation}
where $C$ is the number of Hill-sphere radii on each side of the
planetary core from which it is able to capture planetesimals; 
$C=4$ in our simulations. Once the core mass approaches $M_\mathrm{iso}$,
the $dM_\mathrm{core}/dt$ slows down drastically, and beyond that point,
 gas accretion continues and surpasses the core accretion rate.
The core mass increases to a value of about $\sqrt{2} M_\mathrm{iso}$ 
at crossover, when $M_\mathrm{core} = M_\mathrm{env}$
\citep{pol96}. This phase of relatively slow accretion rates onto
both core and envelope is known as `Phase 2'.

\section{Disk-limited Gas Accretion Rates} \label{sec:acc-rates}

The epoch of rapid gas accretion  in the core-nucleated accretion model
generally begins soon after the envelope mass, $M_{\mathrm{env}}$, 
exceeds the core mass, $M_{\mathrm{core}}$, as can also be shown 
by means of simple thermodynamical arguments \citep{ddl2010}. 
In a proto-solar nebula at $\sim 5\,\mathrm{AU}$, this condition
typically occurs when the planet mass $M_{p}=M_{\mathrm{core}}+M_{\mathrm{env}}$ is between
$\sim 10$ to a few tens of  Earth masses.
After this point, the planet's envelope tends to contract very rapidly, 
limited only by the rate of energy escape at the surface, and  a high rate
of gas accretion is required to maintain the condition $R_p=R_\mathrm{eff}$.
At or about 0.25~M$_\mathrm{Jup}$  this condition can no longer be met,  the
rate is set by the ability of the protoplanetary disk to deliver
gas to the planet, and $R_p$ contracts well within $R_\mathrm{eff}$.

There are various regimes of disk-limited gas accretion \citep[see][]{dl2008}.
For the purpose of this study, we are mainly interested in the high-mass limit
$R_{H}\gtrsim H_{p}$, where $R_{H}=a_{p} \sqrt[3]{M_{p}/\left(3M_{\star}\right)}$
is the Hill radius of the planet and $H_{p}$ is the disk thickness at the planet's 
orbital radius, $a_{p}$. In this regime, disk-limited accretion rates can be
affected by disk-planet gravitational interactions if tidal torques overcome 
viscous torques. Assume that the turbulent (kinematic) viscosity of the
disk at the orbital distance of the planet is given by 
$\nu_{t}=\alpha_{t} H^{2}_{p} \Omega_{p}$, where $\Omega_{p}$ is the
local Keplerian rotation frequency of the disk and $\alpha_t$ is the
viscosity parameter.  Then tidal torques exerted by 
the planet on the disk exceed viscous torques exerted by adjacent disk rings
on each other if
\begin{equation}
\left(\frac{M_{p}}{M_{\star}}\right)^{2} \gtrsim %
3\pi f \alpha_{t} \left(\frac{H_{p}}{a_{p}}\right)^{2} %
\left(\frac{\Delta}{a_{p}}\right)^{3},
\label{eq:tor-con}
\end{equation}
where $\Delta=\max{(H_{p},R_{H})}$ and $f$ is a factor of order unity
\citep[see, e.g.,][and references therein]{ddl2010}.
When the left-hand side of Equation~(\ref{eq:tor-con}) is much greater than the
right-hand side, a gap forms in the disk surface
density along the planet's orbital radius.

We estimated disk-limited accretion rates, $\dot{M}_{p}$, using high 
resolution 3D hydrodynamical simulations of a planet embedded in 
a protoplanetary disk. 
We used an approach along the lines of \citet{dkh2003}. We considered
a disk with a constant aspect ratio of $H_{p}/a_{p}=0.05$ and with the   
parameter $\alpha_{t}$ ranging from $4\times 10^{-4}$ to 
$2\times 10^{-2}$. The unperturbed surface density of the disk is taken to
be a power-law of the distance from the star with exponent $-1/2$.
The planet is kept on a fixed circular orbit and the continuity and 
Navier-Stokes equations (written in terms of linear and angular momenta) 
for the gas are solved in a reference frame co-rotating with the planet.
 
The disk is assumed to be vertically isothermal and, radially, the temperature
drops as the inverse of the distance from the star. At the radial distance $a_{p}$,
the temperature is
$T_{p}=(\mu_d m_{\mathrm{H}}/k_{B}) H^{2}_{p} \Omega^2_{p}$, which
is equal to $53.8\mu_d\,\mathrm{K}$ at $5\,\mathrm{AU}$ from a solar-mass 
star ($\mu_d$ indicates the mean molecular weight of the disk's gas). 
The relatively simple equation of state adopted here (the pressure $p\propto T\rho$,
where $\rho$ is the mass density and the temperature $T$ is a given function
of the orbital distance) allows us to write the fluid equations in a 
non-dimensional form so that the gas accretion rates can be expressed in 
terms of $a^{2}_{p}\Sigma_{p}\Omega_{p}$, where $\Sigma_{p}$ is the
unperturbed gas surface density of the disk at the planet location (i.e., that the disk
would have in the absence of the planet). Furthermore, the planet's mass
enters the calculations only via its ratio to the mass of the star.
We considered values of the ratio $M_{p}/M_{\star}$ up to $0.02$.
The planet's gas accretion rate starts to decline for planet masses 
greater than the value 
for which the inequality in Equation~(\ref{eq:tor-con}) is satisfied. 
This critical  mass is larger within
more  viscous disks.        The equation also suggests that there
is a dependence on the disk thickness, which, however, was not 
explored here. 
We notice that reasonable values of $H_{p}/a_{p}$ for evolved disks,
between $1$ and $10\,\mathrm{AU}$, range 
from $\approx 0.03$ to $\approx 0.05$ \citep[e.g.,][]{dm2012}, affecting the right-hand side of  
Equation~(\ref{eq:tor-con}) by a factor of less than $3$ (for $R_{H}>H_{p}$), 
whereas uncertainties on 
$\alpha_{t}$ are much larger, spanning $2$ orders of magnitude or more.
An unperturbed surface density with a power index different from that adopted 
here ($-1/2$) may also affect the accretion rates. We expect these effects to be small,
especially when tidal torques exerted by the planet drastically modify the surface
density, which is typically the case in the models discussed here.

\begin{figure}[]
\centering%
\resizebox{1.0\linewidth}{!}{%
\includegraphics[]{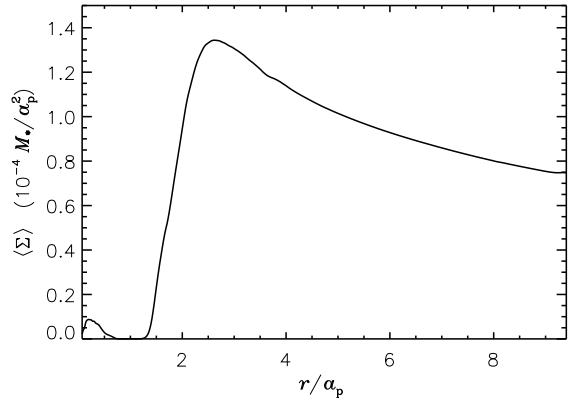}}
\caption{Averaged surface density of gas in a disk, as a function 
of distance from the star, that tidally interacts with a 
planet with   $M_{p}=10^{-2}\,M_{\star}$. The orbital radius of the planet is $a_{p}$.
   The surface density is plotted in scaled units of $M_{\star}/a^{2}_{p}$.
   The aspect ratio of the disk is $H_{p}/a_{p}=0.05$ and the 
               turbulent viscosity parameter is $\alpha_{t}=10^{-2}$.
             }
\label{fig:sigma}
\end{figure}

In the calculations, the disk domain extends in radius as close to the star as 
$0.1\,a_{p}$ (and $0.05\,a_{p}$, in some calculations) and as far as $9.4\,a_{p}$. More vigorous perturbations
exercised by larger mass planets cause  the inner/outer disk radius  to 
decrease/increase with increasing planet-to-star mass ratio.
Figure~\ref{fig:sigma} shows the surface density, averaged in azimuth, for
a case in which $M_{p}=10^{-2}\,M_{\star}$ and $\alpha_{t}=10^{-2}$.
Notice that the low densities in the disk inside the orbit of the planet are
a consequence of tidal torques and planet accretion \citep[see][]{ld2006},
with possibly some impact from the finite radius of the grid inner boundary
($0.05 a_{p}$ in the calculation shown in the figure).
The analysis of \citet{ld2006}, where applicable, suggests that the effects 
of the finite inner grid radius are small.

High resolution in and around the planet's Hill sphere is achieved by means 
of multiple nested grids \citep{dhk2002,dkh2003} centered on the planet's position. 
This methodology allows us to solve the fluid equations (locally around the planet),
and hence to resolve the accretion flow, on length scales of order $0.01\,R_{H}$,
or $\approx 7$ R$_\mathrm{Jup}$ at $5$~AU.

The gas that orbits the planet deep within its gravitational potential is eventually 
accreted. We assume that gas can be accreted within a spherical region of 
radius $0.1\,R_{H}$ (or $0.05\,R_{H}$ in some models), centered on the planet.
The amount of accreted gas is proportional to the amount of gas available
in the region \citep[see][and references therein]{dl2008}. 
In these calculations, accreted gas is removed from the computational domain
but not added to the mass of the planet in order to achieve a stationary 
accretion flow \citep[see][]{lis09}.

\begin{figure}[]
\centering%
\resizebox{\linewidth}{!}{%
\includegraphics[]{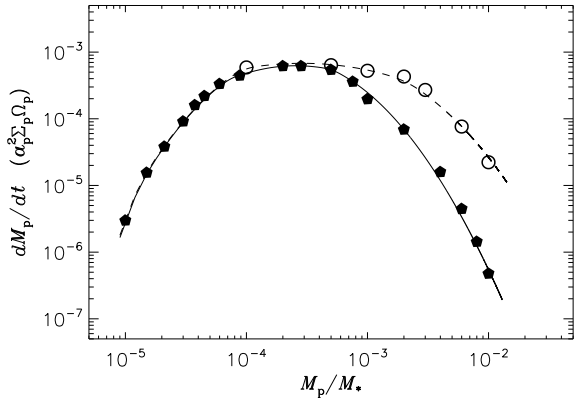}}
\caption{Disk-limited gas accretion rates as a function of the planet-to-star
              mass ratio, $M_{p}/M_{\star}$, for  $H_p/a_p=0.05$ and for
              two values of the disk turbulent 
              viscosity parameter: $\alpha_{t}=4\times 10^{-3}$ and $10^{-2}$.
              Symbols are data obtained from 3D hydrodynamics simulations: 
              filled pentagons/open circles refer to the lower/higher viscosity case.
              The solid and dashed lines represent results from the fitting 
              procedure outlined in the text. In the units of $\dot{M}_{p}$, 
        $\Sigma_{p}$ represents the unperturbed disk gas surface density and 
        $\Omega_{p}$ the Keplerian rotation rate at the planet's orbital
        radius, $a_{p}$.
             }
\label{fig:dmdt}
\end{figure}

We determined an interpolation procedure for the disk-limited gas accretion
rates obtained from calculations,  by performing piece-wise parabolic fits 
(in a logarithmic plane) 
to each $(\dot{M}_{p},M_{p}/M_{\star})$ data set, relative to a given value of the 
turbulent viscosity. Two such fitting curves are shown in Figure~\ref{fig:dmdt}
(explicit expressions are provided in Appendix~\ref{sec:acc-rates-fit}).
Linear interpolations among these curves provide the accretion rate at the 
desired viscosity parameter, $\alpha_{t}$. In doing so, we derived a function 
$\dot{M}_{p}=\dot{M}_{p}(M_{p},M_{\star},a_{p},\Sigma_{p},\alpha_{t})$, which
we employ in our planet formation calculations.
We recall that $\Sigma_{p}$ here represents the disk gas surface density at the
planet's orbital radius, in the absence of the planet. An analytic formula is
available for the accretion rate at the low-mass end \citep{dl2008}.
However, in the formation calculations the use of these curves is not
required until $M_p$ exceeds $\approx 0.25$~M$_\mathrm{Jup}$.

During the disk-limited
gas accretion phase, the solid accretion rate is arbitrarily limited to a
fraction of the value at crossover; the precise value has 
practically no effect on the results.  We don't expect the core-accretion
prescription to be valid at this stage, because most solids in the disk
will not be in the form of planetesimals, and we do not have the
capability to model giant impacts. As the planet reaches within 2\% of
the desired  final mass (e. g. 12~M$_\mathrm{Jup}$) the gas accretion rate, 
already quite low, is smoothly reduced to zero.

\section{Calculations and Results} \label{sec:results}

A recent paper on deuterium burning in objects formed through the
core-accretion scenario \citep{mol12} considered the basic case of
a body  forming at 5.2 AU in a disk around a 1~M$_\odot$ star
with a solid surface density of $\sigma= 10\,\mathrm{g\,cm}^{-2}$ and 
$T_\mathrm{neb}=150$~K. Their study compared results obtained 
by varying the following
parameters: initial entropy of the object after formation (hot start vs. 
cold start), helium abundance, metal abundance, initial deuterium
mass fraction, $\sigma$, which determines the final
 planet core mass, and maximum gas accretion rate.
Their calculations differ from ours in  the phase of rapid gas
accretion, when disk-limited rates apply. They take that rate to be an
arbitrary parameter, while we use the three-dimensional simulations
mentioned above (Section \ref{sec:acc-rates}) to determine it.       
Here we concentrate on cold-start models and consider a somewhat
different set of parameters: stellar mass, formation position of the
planet in the disk,  solid surface density  $\sigma$, 
method of computation of the opacity in the
planetary envelope during the formation phase, 
and protoplanetary disk viscosity parameter $\alpha_t$. The planet's
core mass is determined through the calculation itself, and it depends
on the first three of these quantities.  Note that the final core masses
found in our calculations fall in the range 4.8--31~M$_\oplus$,
while those of \citet{mol12} are higher (30--100~M$_\oplus$).
The formation and evolution
are assumed to take place at a fixed orbital radius. 

\begin{table*}
 \caption{Input Parameters}\label{table:1}
 \centering
% \vspace*{1ex}%
 %\vskip 0.2 in%
 \resizebox{.9\textwidth}{!}{%
 \small
 \begin{tabular}{|l||cccccccc|c|}
 \hline
 Run
 &  M/M$_\odot$ & Distance (AU) & $\sigma$ ($\mathrm{g\,cm}^{-2}$)  & $\rho_{\mathrm{neb}}$ 
 ($\,\mathrm{g\,cm}^{-3}$)
 & $T_{\mathrm{neb}}$ (K)  &  opacity  & $\alpha_t$
 & ${M}_{\rm iso}$ (M$_{\oplus}$)
   \\
 \hline\hline
1A 
 &         1   & 5.2 &10 
 &  $9 \times 10^{-11}$    & 115 &gs
 &  $1.0 \times 10^{-2}$ &$11.6$
  \\
 \hline
1B           
 &         1   & 5.2 &10 
 &  $9 \times 10^{-11}$    & 115 &ngs
 &  $1.0 \times 10^{-2}$ &$11.6$
    \\
 \hline
1C
 &         1   & 5.2 &4 
 &  $3.7 \times 10^{-11}$    & 115 &gs
 &  $1.0 \times 10^{-2}$ &$2.9$
   \\
 \hline
2A         
 &         2   & 9.5 &4
 &  $1.8 \times 10^{-11}$    & 125 &ngs
 &  $1.0 \times 10^{-2}$ &$12.6$
    \\
 \hline
2B
 &         2   & 9.5 &4
 &  $1.8 \times 10^{-11}$    & 125 &ngs
 &  $4.0 \times 10^{-3}$ &$12.6$
    \\
 \hline
2C
 &         2   & 9.5 &6
 &  $2.8 \times 10^{-11}$    & 125 &ngs
 &  $1.0 \times 10^{-2}$ &$23.2$
   \\
 \hline
 \end{tabular}
       }
\end{table*}
%%%%

The parameters for the runs are given in Table~\ref{table:1}. The columns
in the table give, respectively, the run identifier, the mass of the central
star in M$_\odot$, the distance of the planet from the star, the solid
surface density $\sigma$, the density $\rho_\mathrm{neb}$ at 
the surface of the planet during
the earlier phases when this surface connects with the disk, the temperature
$T_\mathrm{neb}$
at the surface during the same phases, the method of opacity calculation 
during the
formation phase---that is, whether it includes the calculation of grain 
settling and coagulation
(gs) or not (ngs)---, the value of the viscosity parameter $\alpha_t$ in the
disk during the phases of disk-limited gas accretion, and the isolation mass
(Equation~\ref{eq:iso}).

Some results for the six runs  are presented in Table~\ref{table:2}.
Each run  is given two lines, the first for a final planet mass that burns
less than half of its deuterium, the second for a nearby mass that
burns more than half. The columns give, respectively, the run
identification, the final planet mass in M$_\mathrm{Jup}$, the total
time to reach the final mass (the formation time), the final core
mass, the central temperature  ($T_\mathrm{c,f}$, at the core/envelope interface) just
after formation, the maximum central temperature during D-burning,
the central density $\rho_\mathrm{c,f}$ just after formation, the planet's radiated
luminosity $L_\mathrm{c,f}$ just after formation, and the mass fraction of deuterium
that remains after 4 Gyr of evolution,  in units of the initial D mass
fraction of $4 \times 10^{-5}$.

\subsection{Results for 1~M$_\odot$}

Run 1A, with standard parameters of 1~M$_\odot$, 5.2~AU, and 
$\sigma=10\,\mathrm{g\,cm}^{-2}$
was originally calculated by \citet{mov10} through most of the formation
phase, including the detailed calculation of grain opacity (their run 
$\sigma10)$.  Their run, whose characteristics are listed in that paper,
 ended at the beginning of disk-limited gas accretion, with a core mass of
16.8~M$_\oplus$  and an envelope mass of 56.8~M$_\oplus$
at a total elapsed time of 1 Myr. In this work, the run was continued
through the disk-limited phase with $\alpha_t = 10^{-2}$ 
(Section \ref{sec:acc-rates})
up to the mass range required for deuterium burning. The maximum gas
accretion rate was $2.5 \times 10^{-1}$~M$_\oplus$ yr$^{-1}$ at a total
planet mass of 96~M$_\oplus$, declining to $10^{-2}$~M$_\oplus$~yr$^{-1}$
at 10~M$_\mathrm{Jup}$. For several different
masses in that range, the accretion was terminated, the opacity was reset
in the surface layers to the values given by \citet{fre08}, and the
evolution was followed at constant mass up to Gyr times. The runs were
terminated when deuterium burning ceased, and the mass $M_{50}$, where 50\%
of the original deuterium had been burned, was determined. In the Run  1A,
the total formation time at $M_{50}$, up to termination of accretion, 
was 1.2 Myr, well within the lifetime of protoplanetary disks.

\begin{figure}[]
\centering%
\resizebox{\linewidth}{!}{%
\includegraphics[]{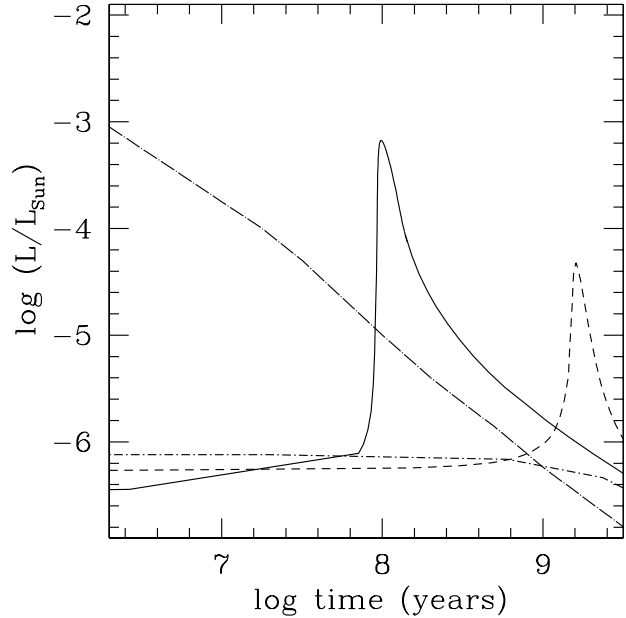}}
\caption{Luminosity (in solar units)  as a function of time for Run 1A during the
post-formation deuterium-burning phase for three different planet masses.
{\it Solid curve:} 16~M$_\mathrm{Jup}$, {\it dashed curve:} 
12.5~M$_\mathrm{Jup}$, {\it short-dash dot curve:} 12~M$_\mathrm{Jup}$.
The {\it long-dash dot curve} shows the results for a hot-start model
of  10~M$_\mathrm{Jup}$ \citep{bar03}.
             }
\label{fig:3}
\end{figure}

\begin{figure}[]
\centering%
\resizebox{\linewidth}{!}{%
\includegraphics[]{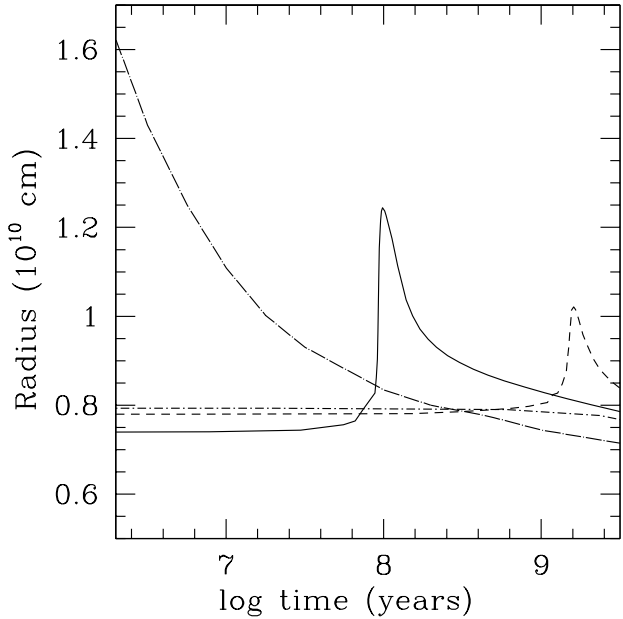}}
\caption{Radius  (in $10^{10}$~cm)  as a function of time for Run 1A during the
post-formation deuterium-burning phase for three different planet masses.
{\it Solid curve:} 16~M$_\mathrm{Jup}$, {\it dashed curve:} 
12.5~M$_\mathrm{Jup}$, {\it short-dash dot curve:} 12~M$_\mathrm{Jup}$.
The {\it long-dash dot curve} shows the results for a hot-start model
of  10~M$_\mathrm{Jup}$ \citep{bar03}.
             }
\label{fig:rad}
\end{figure}

The planetary luminosity as a function of time for three  different final
masses  in Run 1A is shown in Figure~\ref{fig:3}, where it is compared
with that typically obtained in a `hot-start' model. In the case of 
16~M$_\mathrm{Jup}$, just after formation the central temperature 
$T_\mathrm{c,f}= 2.8 \times 10^5$~K, too low for substantial burning  
on a short time scale, even though 
$\rho_\mathrm{c,f}=80\,\mathrm{g\,cm}^{-3}$. Under these conditions the 
screening correction factor to the nuclear reaction rate is high, about  88. 
Consequently, deuterium burning can take place at relatively low
temperatures compared to those ($\approx 10^6$~K) where deuterium
burns in solar-mass stars.  The central
temperature $T_c$ as a function of time (at the core/envelope interface) 
gradually increases as a result of slow deuterium burning and is
accompanied by a slight increase in radius.  When  $T_c$ reaches
$3.2 \times 10^5$~K, 
a rapid increase in burning occurs, leading to a peak in
luminosity at about $10^8$ years.  At the peak about 60\% of the 
deuterium has burned,  and  $T_c$ is near its maximum
of $5.1 \times 10^5$~K.  At the same time the radius has
increased from $7.4 \times 10^9$~cm to $1.25 \times 10^{10}$~cm; 
then it contracts
again after the luminosity peak. At the end of the evolution essentially
all the deuterium has burned.   A similar process, involving a
rapid increase in deuterium burning in the context of a  slowly accreting
brown dwarf, was studied by \citet{sal92}; he denotes the
event a `deuterium flash'. The radii for the three masses, as well as
for the hot-start case, are shown in Figure~\ref{fig:rad}. 
The general result that
cold-start models result in a radius increase during deuterium
burning agrees with the previous results of \citet{mol12}.

  In the case of 12.5~M$_\mathrm{Jup}$, right after formation the central
temperature is lower, only $2.6 \times 10^5$~K, with a central density
of $52\,\mathrm{g\,cm}^{-3}$ and a screening factor of 70. It takes almost $10^9$ years
for rapid deuterium burning to start, at $T_c= 2.8 \times 10^5$~K, with the burning occurring on a much longer time scale than in the
case of 16~M$_\mathrm{Jup}$.
 Eventually about 98\% of the deuterium is burned, and the
luminosity peak, which is somewhat lower, is shifted to later times. At the 
peak, about half of the D has burned, and this point is also close to the
maxima in radius and $T_c$.               
In the case of 12~M$_\mathrm{Jup}$ only 6\% of the deuterium is burned, 
and no peak in luminosity appears. At $M_{50}$ itself, the peak
involves only a factor 2 increase in luminosity.  In the peaks, the total energy
$\int L dt$ is found to agree closely with the  total energy available
from D-burning, given by the quantity $Q_\mathrm{dp}/m_d
\times M_p \times X_d f_d$, where $Q_\mathrm{dp}$, the energy production per reaction is
expressed in ergs, $m_d$ is the mass of a deuterium atom,  $M_p$ is the
planet mass, $X_d$ is the initial mass fraction of deuterium, and $f_d$ is
the fraction of the initial D that burned. The total energy is about
$2.5 \times 10^{45}$ ergs for  the 12.5~M$_\mathrm{Jup}$ case.

Results for two runs whose final
masses closely bracket $M_{50}$ are shown in Table~\ref{table:2}. 
The main result of
this case is that $M_{50}=12.37$~M$_\mathrm{Jup}$ with a heavy-element core
mass of 16.8~M$_\oplus$.  By way of comparison, a cold
start model with a core, calculated by \citet{mol12}  
with about the same basic parameters 
(1~M$_\odot$, $\sigma=10\,\mathrm{g\,cm}^{-2}$, $a_p=5.2$ AU), 
with a similar  helium mass fraction of 28\%, 
but with some differences in assumptions and computational
procedure, gives $M_{50}=12.6$~M$_\mathrm{Jup}$. 

The maximum $T_c$   at the core/envelope
interface  for $M_{50}$ in this case is close to $3.2 \times 10^5$~K, a very sensitive
function of mass. Whether significant
D-burning occurs depends sensitively on this temperature. If it reaches, say
$2.5 \times 10^5$~K, practically no D is burned for the corresponding mass
of 12.0~M$_\mathrm{Jup}$.  If it reaches
$4.0 \times 10^5$~K, practically  all (98\%) of the  D is burned for the 
corresponding mass of 12.5~M$_\mathrm{Jup}$. Once the threshold is
reached, energy deposition from burning increases the temperature,
which increases the reaction rate, as it is proportional to $T^{12}$.
The resulting expansion leads to a near thermal equilibrium, with the
energy produced from D-burning matched closely by the total radiated luminosity.

Run 1B differs from 1A only with respect to the calculation of the opacity
resulting from grains
in the protoplanetary envelope during the formation phase. As mentioned
above, in Run 1A this opacity is obtained through detailed consideration
of grain settling and coagulation \citep{mov10}. In 1B  a table of
interstellar grain
opacities is used, reduced by a factor of about 50. The characteristics of
this run, up to a mass of about 1~M$_\mathrm{Jup}$, are very similar
to those listed for Run 1sG in \citet{lis09}. The crossover mass is
16.16~M$_\oplus$, the crossover time is 2.31 Myr, and the onset of disk-limited
rapid gas accretion  occurs at a core mass of 16.8~M$_\oplus$ and a time
of 2.41 Myr. Note that the evolution time up to this point is 2.4 times
longer than in Run 1A.  Note also that the core mass is the same as in Run 1A;
the substantial difference in opacity, which can be up to two orders of 
magnitude in certain ($\rho,T$) regions, has practically no effect on
the core mass.  

 Here, the disk-limited accretion rates are used 
to continue the evolution up to the D-burning mass range. The luminosity
as a function of time up to the end of accretion is shown in 
Figure~\ref{fig:lum}. The results for D-burning after that time
show that $M_{50} = 12.20$~M$_\mathrm{Jup}$, not significantly different
from the results of Run 1A. As Table~\ref{table:2} shows, $T_\mathrm{c,f}$  
in Run 1B, at the same final mass, is  slightly higher than
that in Run 1A, just after formation. Correspondingly, $\rho_\mathrm{c,f}$ 
is slightly    lower. These small differences indicate a slightly
higher entropy for  1B after formation, as indicated by the slightly
higher luminosity at this point. The increased envelope opacity
in 1B as compared with 1A results in slower heat loss and tends to
keep internal temperatures higher. However this effect is almost
compensated by the fact that the formation time is more than 
twice as long in  1B.  Even the slight increase in $T_\mathrm{c,f}$ in 1B as compared
with 1A allows $M_{50}$ to be pushed to a slightly lower mass.

Run 1C differs from Run 1A in that $\sigma$ is reduced to $4\,\mathrm{g\,cm}^{-2}$,
a value only slightly greater than that in a minimum-mass solar nebula
\citep{wei77}. 
Grain settling and coagulation are included in the
opacity calculation. The earlier portions of this run, up to the onset
of disk-limited gas accretion, are described in \citet{mov10}, their
run $\sigma4$.  The time to reach this point, 3.5 Myr, is considerably
longer than in Run 1A,  first,  because the core accretion rate is considerably
lower, and second, because the lower isolation mass results in reduced
luminosity and reduced gas accretion rate during Phase 2 \citep{pol96}.
The crossover mass is 4.09~M$_\oplus$, and the core mass at the time
of onset of disk-limited accretion is 4.74~M$_\oplus$.

The calculations were continued up to the point where gas accretion
terminated, at which point the core mass was 4.8~M$_\oplus$. The total time 
to reach 12~M$_\mathrm{Jup}$ was about 4.1~Myr, and to 14~M$_\mathrm{Jup}$,
about 4.5 Myr.  The peak disk-limited accretion rate was
$1.0 \times 10^{-1}$~M$_\oplus$ yr$^{-1}$ at 0.3~M$_\mathrm{Jup}$, a factor
of 2.5 lower than in Run 1A because of the reduction in $\Sigma_{p}$ by the
same factor.   By the time the total mass
was 5~M$_\mathrm{Jup}$ the rate was down to 
$1.8 \times 10^{-2}$~M$_\oplus$~yr$^{-1}$, and at 10~M$_\mathrm{Jup}$ it had further
declined to  $4.3 \times 10^{-3}$~M$_\oplus$~yr$^{-1}$. Much of the time
during the disk-limited accretion phase was spent in accreting the last
1--2~M$_\mathrm{Jup}$ to reach the D-burning point. The luminosity as
a function of time for this run, up to the end of accretion,  
is shown in Figure~\ref{fig:lum}.

The luminosity versus time plots for Run 1C during the D-burning phase
look similar to those
for 1A, except in this case M$_{50}$ noticeably increases to
13.55~M$_\mathrm{Jup}$. The reduced core mass in 1C (4.8~M$_\oplus$)
as compared to that in 1A (16.8~M$_\oplus$) is clearly associated with
the difference, in agreement with the results of  \citet{mol12}.
In our calculations,
the core equation of state gives a core radius of $3.8 \times 10^8$~cm
for the Run 1C core of mass 4.8~M$_\oplus$ when the total mass is 12
M$_\mathrm{Jup}$.
For the core of 16.8~M$_\oplus$ in Run 1A, at the same total mass, the
radius is $6.0 \times 10^8$~cm. Thus, at the core boundary,
the gravitational potential is more negative, and the gravity is about
40\% greater in 1A than in 1C.
The calculated values of $T_\mathrm{c,f}$ are
$\approx 2.6 \times 10^5$~K and $\approx 2.1 \times 10^5$~K in Runs
1A and 1C, respectively.

It follows from the equation of hydrostatic equilibrium  \citep{mol12}
that in a convective envelope the adiabatic temperature gradient at the
interface should
be proportional to the core gravity, so a higher
gravity most probably gives a higher temperature. However, this
statement is inconclusive. We calculated static
models for  a planet  of 12~M$_\mathrm{Jup}$, all with the  envelope entropy of Run 1C, with
core masses ranging from 0 to 15~M$_\oplus$. We found practically no
difference in  $T_c$ as a function of $M_\mathrm{core}$, with
$T_c$ {\it decreasing} by less than 1\% when the core mass
increases from 0 to 15 $M_\oplus$.

The real source of the difference in $T_\mathrm{c,f}$ between Runs 1A
and 1C is the entropy in the envelope.
The lower $T_\mathrm{c,f}$ and higher $\rho_\mathrm{c,f}$
for 1C as compared with 1A indicate a lower entropy, which is consistent
with the fact that the luminosity just after formation is lower by more
than a factor 2 in 1C (Table~\ref{table:2}). The values of entropy just
after formation for  a planet of 12~M$_\mathrm{Jup}$ in  Runs
1A and 1C are, respectively, 8.02 and 7.52 $k_B$ per baryon.  The
entropy is determined
through the physical processes that occur during the entire formation
phase; for example, the formation time for Run 1C is almost 4 times
longer than that for 1A, and the same opacities were used,
 which suggests a lower entropy.
Thus there exists a qualitative understanding of
the relation between core mass and $T_\mathrm{c,f}$, but a quantitative
theory, apart from the numerical simulations, is quite difficult.

In Run  1C, $T_\mathrm{c,f}=2.1 \times 10^5$~K is the maximum reached
for a final mass of
12~M$_\mathrm{Jup}$, and it is insufficient for D-burning.  In the case
of Run 1A, the corresponding $T_\mathrm{c,f}$ is much closer to the
threshold
required for burning. Thus the planet with the higher $M_\mathrm{core}$
is able to produce significant D-burning at a lower total mass.
As Table~\ref{table:2} shows, in the mass range for
Run 1C where D-burning begins, just above 13.5~M$_\mathrm{Jup}$,
$T_\mathrm{c,f}$ is somewhat less ($2.3 \times 10^5$~K) than in the
corresponding mass range for Run 1A. However, to compensate,
$\rho_\mathrm{c,f}$ is higher, about $65\,\mathrm{g\,cm}^{-3}$,  and the screening
factor at the center has increased to 160.
Again, the lower entropy at formation for 1C, as compared with 1A,
a result of   various processes associated  with the accretion of
core and envelope,  leads to a higher $M_{50}$.

\begin{table*}
 \caption{Selected Results}\label{table:2}
 \centering
% \vspace*{1ex}%
%  \vskip 0.2 in%
 \resizebox{0.9\textwidth}{!}{%
 \small
 \begin{tabular}{|l||cccccccc|c|}
 \hline
 Run
 &  $M_\mathrm{final}$/M$_\mathrm{Jup}$ & $t_\mathrm{form}$ (yr) & 
$M_\mathrm{core}$/M$_\oplus$ & $T_\mathrm{c,f}$ (K)
 & $T_\mathrm{max}$ (K)  &  $\rho_\mathrm{c,f}$  ($\mathrm{g\,cm}^{-3}$)  & 
 log ($L_\mathrm{f}/\mathrm{L}_\odot$) & $D_\mathrm{final}$/$D_\mathrm{init}$
   \\
 \hline\hline
1A 
 &   12.26     & $1.19 \times 10^6$ &16.8 
 &  $2.60 \times 10^5$    & $2.60 \times 10^5$ & 49.7 
 &   -6.22 &$0.895 $
  \\
 \hline
1A 
 &   12.48     & $1.20 \times 10^6$ &16.8 
 &  $2.62 \times 10^5$    & $3.54 \times 10^5$ & 51.4 
 &   -6.23 &$0.164$
  \\
 \hline
1B           
 &  12.14     & $2.67 \times 10^6$ &16.8 
 &  $2.69 \times 10^5$    & $2.69 \times 10^5$ &47.0
 &  -5.84 &$0.860$
    \\
 \hline
1B           
 &  12.26     & $2.68 \times 10^6$ &16.8 
 &  $2.72 \times 10^5$    & $3.18 \times 10^5$ &47.7
 &  -5.91 &$0.328$
    \\
 \hline
1C
 &     13.5    & $4.37 \times 10^6$ &4.83 
 &  $2.31 \times 10^5$    & $2.31 \times 10^5$ &65.1
 &  -6.57  &$0.938$
   \\
 \hline
1C
 &     13.6    & $4.39 \times 10^6$ &4.83 
 &  $2.33 \times 10^5$    & $4.27 \times 10^5$ &65.9
 &  -6.57  &$0.292$
   \\
 \hline
2A         
 &     11.9    & $2.14 \times 10^6$ &18.7
 &  $2.73 \times 10^5$    & $2.81 \times 10^5$ &41.4
 &  -5.64 &$0.767$ 
    \\
 \hline
2A         
 &     12.0    & $2.14 \times 10^6$ &18.7
 &  $2.76 \times 10^5$    & $3.06 \times 10^5$ &41.8
 &  -5.63 &$0.435$
    \\
 \hline
2B
 &   12.0      & $3.23 \times 10^6$ &18.8
 &  $2.78 \times 10^5$    & $2.87 \times 10^5$ &44.6
 &  -5.73  &$0.582$
    \\
 \hline
2B
 &   12.1      & $3.26 \times 10^6$ &18.8
 &  $2.80 \times 10^5$    & $3.28 \times 10^5$ &45.1
 &  -5.72  &$0.320$
    \\
 \hline
2C
 &        11.6 & $8.75 \times 10^5$  &31.0
 &  $3.37 \times 10^5$    & $3.46 \times 10^5$ &27.9 
 &  -4.80 &$0.670$
   \\
 \hline
2C
 &        11.7 & $8.75 \times 10^5$  &31.0
 &  $3.39 \times 10^5$    & $3.56 \times 10^5$ &28.3 
 &  -4.80 &$0.390$
   \\
 \hline
 \end{tabular}
       }
\end{table*}
%%%%

\subsection{Results for 2~M$_\odot$}

\begin{figure}[]
\centering%
\resizebox{\linewidth}{!}{%
\includegraphics[]{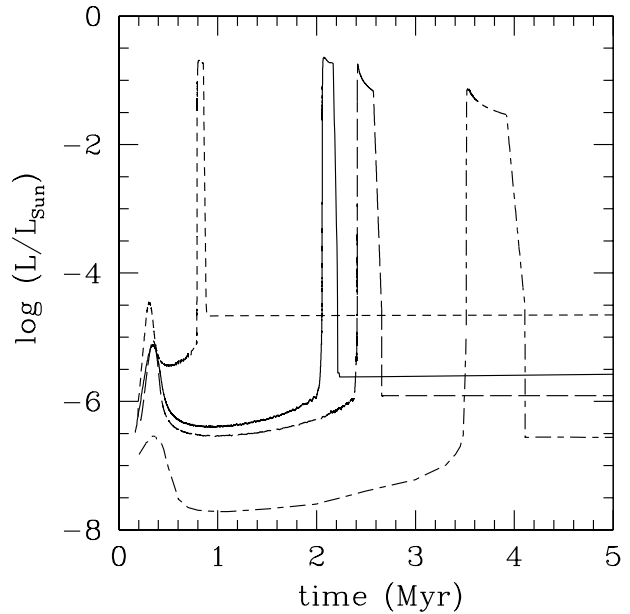}}
\caption{Luminosity (in solar units)  as a function of time for Run 1B 
({\it long-dashed curve}), Run 1C ({\it short-dash long-dash curve}), Run 2A 
({\it solid curve}), and  Run 2C  ({\it short-dashed curve}) during the
formation phase. 
             }
\label{fig:lum}
\end{figure}

\begin{figure}[]
\centering%
\resizebox{\linewidth}{!}{%
\includegraphics[]{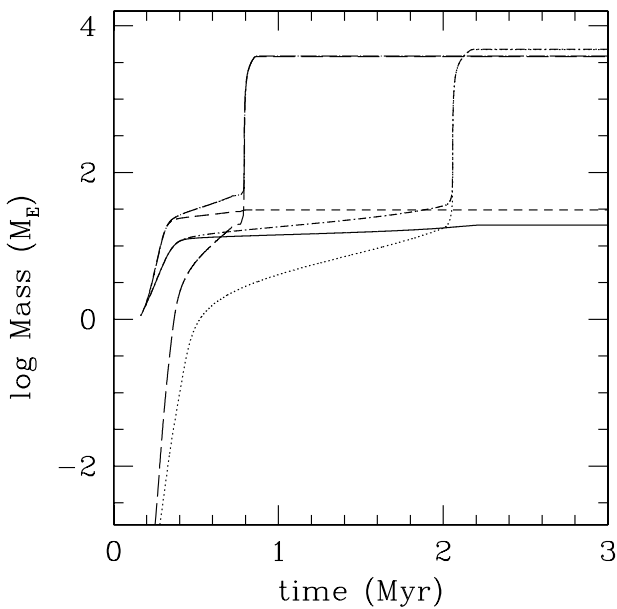}}
\caption{Planet mass (in M$_\oplus$)  as a function of time for Run 2A 
 and Run 2C  during the
formation phase. The final masses are 15~M$_\mathrm{Jup}$
and 12~M$_\mathrm{Jup}$, respectively.
For Run 2A, the {\it solid curve} gives the core mass, the {\it dotted
curve} the envelope mass, and the {\it short-dash dot curve} the total
mass. For Run 2C, the {\it short-dashed curve} gives the core mass, 
the {\it long-dashed curve} the envelope mass, and the {\it long-dash dot 
curve} the total mass.
}
\label{fig:mass}
\end{figure}

The formation phases of Runs 2A and 2C, for a central star of 2~M$_\odot$, 
are illustrated in Figure
\ref{fig:lum}, which gives the luminosity as a function of time, 
 and  Figure~\ref{fig:mass},  which  
 gives the core mass, envelope mass, and total mass as a
function of time. 
Run 2A differs from 1A in that the planet is placed 9.5 AU away from a
star of 2~M$_\odot$, in a disk with $\sigma=4\,\mathrm{g\,cm}^{-2}$. In a minimum
mass solar nebula, scaled to the mass of this star, the corresponding 
value would be $2\,\mathrm{g\,cm}^{-2}$.

The isolation mass, however is quite similar to
that in 1A, 12.6 rather than 11.6~M$_\oplus$. 
 The opacity during the
formation phase  of 2A is taken from a table of interstellar grain opacities,
reduced by a factor of 50, as in Run 1B. However, the comparison between
1A and 1B showed that these opacities have little effect on $M_{50}$.
The formation time is
longer in 2A than in 1A because of the longer dynamical time at the
larger distance, the reduced solid surface density, and the somewhat
higher envelope opacity. However, these effects are partially compensated for
by the smaller planetesimal size (50 km in 2A; 100 km in 1A), which
increases the capture cross section $\pi R^2_\mathrm{capt}$,  and by the increased
gravitational focussing factor $F_g$ at the larger distance. 

The first luminosity peak for Run 2A (Figure~\ref{fig:lum}) occurs at $t=3.54 \times 10^5$ 
yr, with log $L$/L$_\odot = -5.14$, with $M_\mathrm{core}=9.3$~M$_\oplus$, with
$M_\mathrm{env}= 0.024$~M$_\oplus$, and  
 with $\dot M_\mathrm{core} = 6.67 \times 10^{-5}$~M$_\odot$ yr$^{-1}$.
This peak corresponds to the maximum in the accretion rate of solids 
onto the core.  The crossover
mass (Figure~\ref{fig:mass})  of 17.6~M$_\oplus$ is reached in 
$2.0 \times 10^6$ years.
 The second, much higher luminosity peak at
 $2.07 \times 10^6$ yr corresponds to the phase of rapid gas accretion up to
a final mass of 15~M$_\mathrm{Jup}$. At that time 
the maximum gas accretion rate is $2.2 \times 10^{-1}$~M$_\oplus$~yr$^{-1}$
and  $M_p =  0.47$~M$_\mathrm{Jup}$.
Formation is complete, up to 15~M$_\mathrm{Jup}$,
in $2.2 \times 10^6$~yr.   The final 
$M_\mathrm{core} = 18.7$~M$_\oplus$ is slightly higher than in  Run 1A.

\begin{figure}[]
\centering%
\resizebox{\linewidth}{!}{%
\includegraphics[]{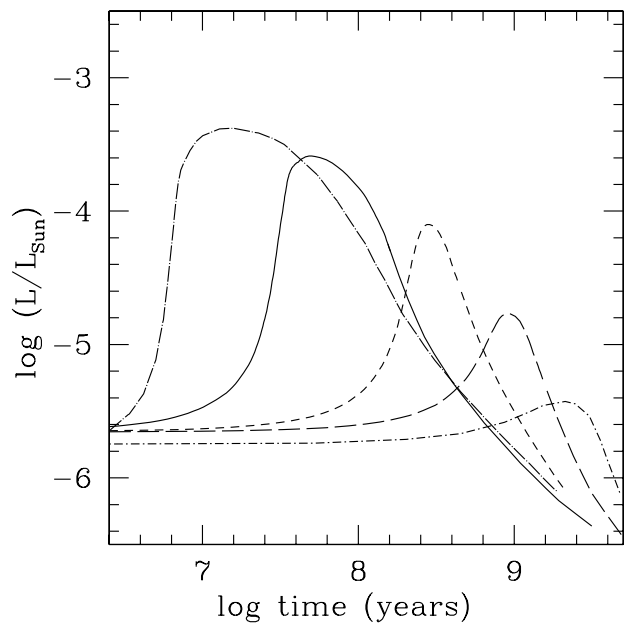}}
\caption{Luminosity (in solar units) as a function of time for Run 2A during the
post-formation deuterium-burning phase for five different planet masses.
{\it Long-dash dot curve}: 16~M$_\mathrm{Jup}$, 
{\it solid curve:} 15~M$_\mathrm{Jup}$, {\it short-dashed curve:} 
13.5~M$_\mathrm{Jup}$, {\it long-dashed curve:} 13~M$_\mathrm{Jup}$,
{\it short-dash dot curve:} 12~M$_\mathrm{Jup}$.
             }
\label{fig:4}
\end{figure}

The luminosity as a function of time during the later deuterium-burning
phase is shown for five different final
masses in Run 2A in Figure~\ref{fig:4}. In the cases of  16 and 
15~M$_\mathrm{Jup}$,
practically all ($>99$\%) of the deuterium is burned; 
in the case of 13.5~M$_\mathrm{Jup}$,
about  92\% is burned; for  13.0~M$_\mathrm{Jup}$, 75\% is burned, and 
for  12.0~M$_\mathrm{Jup}$, just over 50\% is burned. Thus the value
for M$_{50} \approx 11.95$~M$_\mathrm{Jup}$ is very close to the values
obtained for Runs 1A/1B despite substantial differences in assumptions and
initial conditions. 
 As discussed in the comparison between Runs 1A and 1C,
the somewhat larger $M_\mathrm{core}$ in 2A as compared to 1A is the 
main reason for the
slightly lower $M_{50}$ in 2A.  After formation, 2A has a slightly higher
$T_\mathrm{c,f}$ than  1A and slightly lower $\rho_\mathrm{c,f}$, leading to
a slightly higher entropy. 

Comparing the luminosity curves for a mass of 16~M$_\mathrm{Jup}$ in
Figures \ref{fig:3} and \ref{fig:4}, they look very different but in fact
they are consistent. In Run 2A (Figure~\ref{fig:4}) the higher $T_\mathrm{c,f}$  
(because of the somewhat higher $M_\mathrm{core}$) allows D-burning to
start earlier than in Run 1A, and the value of $L$ at the starting point 
is a factor of 4 higher. In fact the full widths of the two curves are quite
similar, the peak values agree to better than a factor 2, and the
integrated luminosities over time of the two curves agree to within 10\%.

Run 2B has exactly the same parameters as 2A except that $\alpha_t$         
is reduced by a factor 2.5, which affects the gas accretion rates during
the disk-limited phase. Thus the formation time in 2B turns
out to be a factor of 1.5 longer at $3.2 \times 10^6$ years, but still
within the range of observed disk lifetimes. Table~\ref{table:2} shows
that for final mass 12~M$_\mathrm{Jup}$ the fractions of deuterium
burned are in agreement for runs 2A and 2B, within the uncertainties of the
calculations. Thus $\alpha_t$ has practically no effect on $M_{50}$
in this case. Run 2B has a slightly lower entropy than 2A at 
12~M$_\mathrm{Jup}$, 8.2 $k_B$ per baryon versus 8.25, 
and therefore a slightly higher $M_{50}$. Thus it appears that the longer     
time during disk-limited accretion in Run 2B has only a 
weak effect on both $M_{50}$ and the entropy, at the same $M_\mathrm{core}$.

Run 2C has the same parameters as 2A except that the solid surface
density $\sigma$ is increased by a factor 1.5 to $6\,\mathrm{g\,cm}^{-2}$. The
first luminosity peak (Figure~\ref{fig:lum})
occurs at $t=3.07 \times 10^5$ years with log $L$/L$_\odot = -4.45$ at 
$M_\mathrm{core}=15.6$~M$_\oplus$. The crossover mass (Figure~\ref{fig:mass}) 
is reached at $t=7.88 \times 10^5$ yr with a value of 30.7~M$_\oplus$.
The maximum luminosity in the second peak is above log $L$/L$_\odot = -1$,
at $t = 7.915 \times 10^5$ years and a total mass of 0.62~M$_\mathrm{Jup}$.
The higher $\sigma$ with respect to Run 2A
results in a markedly higher $M_\mathrm{core}= 31$~M$_\oplus$ and 
a markedly shorter formation time ($8.75 \times 10^5$ yr at $M_{50}$).
Despite these relatively large differences, the value for $M_{50}$ in
2C is only 2.5\% smaller than in 2A. At the end of the formation phase,
central temperatures are higher and central densities are lower in 2C
as compared with 2A. Also, the screening factor is only 14 in 2C compared
with 41 for 2A.  The entropy  at formation, for a final mass of 
12~M$_\mathrm{Jup}$, is higher in 2C,  9.08 
$k_B$/baryon as compared with 8.25, corresponding to a 
higher luminosity at that point. The slope in the ($M_\mathrm{core},
~M_{50}$) diagram between  $M_\mathrm{core}$=  18.7 and 31~M$_\oplus$
is -0.024, a result which differs somewhat from that of
\citet{mol12}. They obtain a slope (in the same units) of -0.01,
although for a different core mass range, 30 to 100~M$_\oplus$.

\begin{figure}[]
\centering%
\resizebox{\linewidth}{!}{%
\includegraphics[]{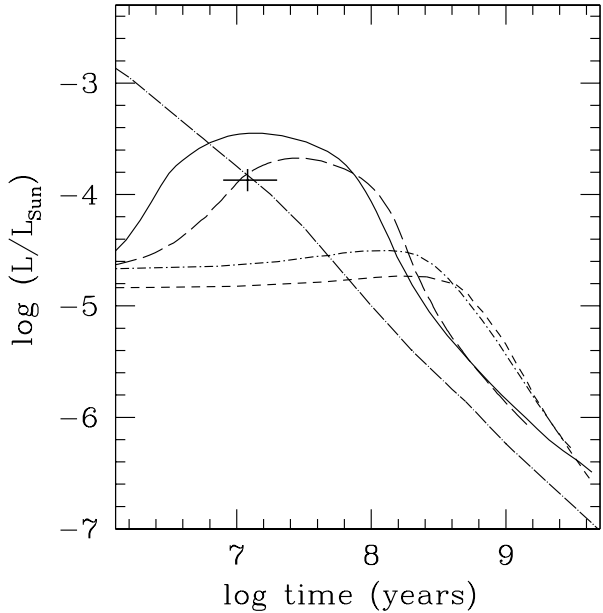}}
\caption{Luminosity (in solar units)  as a function of time for Run 2C during the
post-formation deuterium-burning phase for four  different planet masses.
{\it Solid curve:} 15~M$_\mathrm{Jup}$, {\it long-dashed curve:} 
13.7~M$_\mathrm{Jup}$, {\it short-dash dot curve:} 12~M$_\mathrm{Jup}$,
{\it short-dash  curve:} 11.7~M$_\mathrm{Jup}$. The {\it long-dash dot curve}
shows   a hot-start model for 10~M$_\mathrm{Jup}$ \citep{bar03}. The cross gives the
position and error bars for the companion to Beta Pic \citep{bon13}.
             }
\label{fig:5}
\end{figure}

Plots of luminosity versus time are shown in Figure~\ref{fig:5} 
for several different masses in Run 2C. As in Figure~\ref{fig:4} the
higher masses give higher peak luminosity at earlier times than the
lower masses, and at $M_{50}$ there is only a very small     peak.
The  $L(t)$ curve for 15~M$_\mathrm{Jup}$  starts at a higher value and
reaches a maximum sooner than for the same mass in Run 2A, because of the
higher internal temperature, but the value of log $L$ at the peak is
about the same.
At 13.7 and 15~M$_\mathrm{Jup}$ practically all of the initial D is burned.
 At 12~M$_\mathrm{Jup}$, 72\% is burned, while at 11.7~M$_\mathrm{Jup}$,
which is very close to $M_{50}$, 61\% is burned.

\begin{figure}[]
\centering%
\resizebox{\linewidth}{!}{%
\includegraphics[]{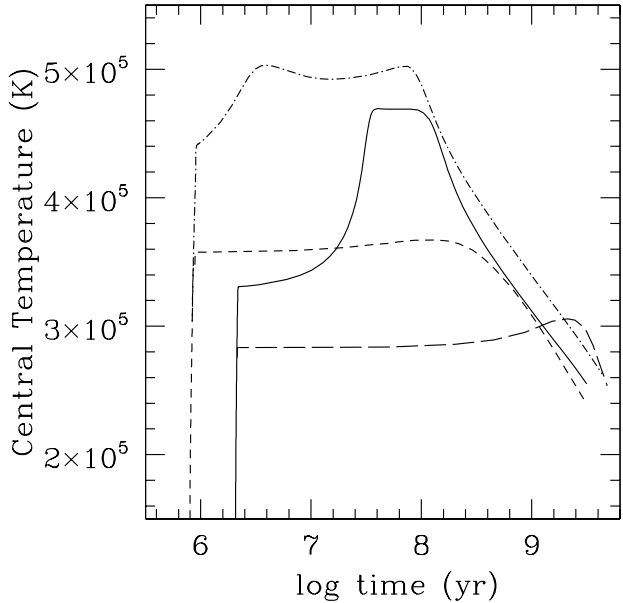}}
\caption{Central temperature $T_c$ (at the core/envelope interface) as a function of time for  four cases. 
{\it Solid curve}: Run 2A (15~M$_\mathrm{Jup}$),
{\it long-dashed curve}: Run 2A (12~M$_\mathrm{Jup}$),
{\it dash-dot curve}: Run 2C (15~M$_\mathrm{Jup}$),
{\it short-dashed curve}: Run 2C (12~M$_\mathrm{Jup}$).
 The phases of rapid gas accretion and final phases of constant mass with 
deuterium burning are shown. Core masses for Runs 2A and 2C are
18.7 and 31~M$_\oplus$, respectively.             }
\label{fig:tem}
\end{figure}

Plots of $T_c$  versus time, during the deuterium-burning
phases,  are shown
for masses 12 and 15~M$_\mathrm{Jup}$ in Figure~\ref{fig:tem}, 
where they are compared with the results from Run 2A. 
The plot shows the effect of
varying the core mass at fixed total mass, and of varying the total mass
at fixed core mass.  For example, for Run 2A at 15~M$_\mathrm{Jup}$ the
maximum $T_c$ is $4.6 \times 10^5$~K, while for 12~M$_\mathrm{Jup}$ it is
only $3.06 \times 10^5$~K and is reached at a much later time. 
The vertical portions of these curves show the effect
of rapid gas accretion from about 1~M$_\mathrm{Jup}$ to the final mass
of either 12 or 15~M$_\mathrm{Jup}$. The two nearly horizontal curves
are for a total mass of 12~M$_\mathrm{Jup}$ and core masses of 18.7 (lower; Run 2A)
and 31~M$_\oplus$ (upper; Run 2C). The higher core mass results in a higher
temperature by a factor of about 1.3. In the case of the 31~M$_\oplus$
core, about 75\% of the deuterium is burned; in the 18.7~M$_\oplus$  core,
a little over 50\%. Note that the D-burning occurs late in the evolution,
where small peaks in the temperature are seen. The remaining two curves
correspond to a total mass of 15~M$_\mathrm{Jup}$, with the same two
core masses just mentioned. The D-burning occurs earlier than in the case
of the lower total mass, and the higher core mass again gives a higher
maximum $T_c$.     In both cases for 15~M$_\mathrm{Jup}$ practically
all the D is burned, and the residual mass fraction is smaller ($8.9 \times
10^{-11}$) for the higher core mass as compared to the lower ($2.6 \times
10^{-9}$).

\begin{figure}[]
\centering%
\resizebox{\linewidth}{!}{%
\includegraphics[bb=110 144 592 718,clip]{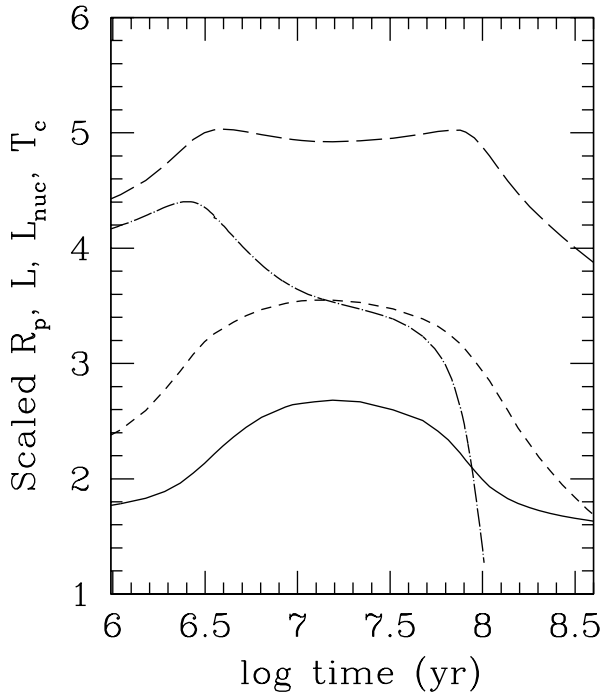}}
\caption{Detail of the deuterium-burning phase for Run 2C, 
15~M$_\mathrm{Jup}$. 
{\it Solid curve}: outer radius $R_p$ as a function of time, in units
of $5 \times 10^9$~cm;
{\it long-dashed curve}: central temperature $T_c$, at the core/envelope interface,
as a function of time, in units of $10^5$~K;
{\it dash-dot curve}: nuclear luminosity as a function of time in units of
log ($L_\mathrm{nuc}$/L$_\odot$) +7; 
{\it short-dashed curve}: radiated luminosity $L$ as a function
of time, in the same units as $L_\mathrm{nuc}$.}
\label{fig:llrt}
\end{figure}

 In Run 2C with 15~M$_\mathrm{Jup}$, $T_c$ goes up to about 
 $5 \times 10^5$~K and there are actually two minor peaks. Figure~\ref{fig:llrt} illustrates
in more detail how various quantities vary during this phase.
In this case, with a high $T_\mathrm{c,f}$, nuclear
burning starts very early. During most of the phase, the object is not
in thermal equilibrium.
The first maximum   in $T_c$ occurs when about 25\% of
the D has burned,  close to the time of the maximum in the nuclear burning
luminosity $L_\mathrm{nuc}$. Here $L_\mathrm{nuc}$ is well above the
radiated luminosity $L$, and the extra power goes into expansion, 
resulting 
in slight cooling of the interior. When half the deuterium has burned ($1.3 \times 10^7$ yr), 
there is a maximum in luminosity and radius, corresponding to the slight
minimum in $T_c$. Then contraction along with a slow decrease in nuclear
burning leads to 
slight heating, and the second maximum occurs when 98\% of the D has
burned. This maximum corresponds to the time when $L_\mathrm{nuc}$ 
starts to drop rapidly and to fall well below $L$. Beyond that point, even though contraction 
is occurring, there is
insufficient burning to maintain the high temperature, 
and the object enters
its final cooling phase.  In contrast, in the case of 12~M$_\mathrm{Jup}$,
the main D-burning in Run 2C takes place at practically constant $T_c$, radius, and $L$,
with a slight maximum in $T_c$ of $3.67 \times 10^5$~K at about $10^8$~yr.
In this case the configuration is close to thermal equilibrium through most
of the D-burning phase.

\subsection{Comparison with beta Pictoris b}

The cross in Figure~\ref{fig:5}  gives the approximate location of the directly imaged
companion \citep{lag09} to the well-known star Beta Pictoris. 
That star, according to \url{http://exoplanet.eu},  has
a mass of about 1.8~M$_\odot$ and an age of 12 ($+8$,$-4$)~Myr. 
The planet is located between 8 and 15 AU from the star \citep{lag10};
thus an approximate comparison can
be made with these calculations. Beta Pic b's position in the
(log $L$, $t$) diagram is plotted in \citet{mar10} where it is shown
to fall on a theoretical track with mass 10~M$_\mathrm{Jup}$ as 
calculated from a `hot start' by \citet{bar03}. In  \url{http://exoplanet.eu} that
mass is given as 8 ($+5$,$-2$)~M$_\mathrm{Jup}$.  The surface temperature
$T_\mathrm{eff}$ has been estimated from observed near infrared colors
\citep{bon11,qua10} at 1700~K, with considerable uncertainty ($\approx 300$~K).
Further infrared and astrometric  observations \citep{bon13} are essentially in agreement,
giving log ($L$/L$_\odot) = -3.87 \pm 0.08$, $T_\mathrm{eff} = 1700 \pm 100$~K, 
$a_p=8-10$~AU, and `hot-start' masses in the range 7--13~M$_\mathrm{Jup}$.
The bolometric luminosity found by \citet{marl} is in agreement with the
above value, and they find `hot-start' masses in the range 7--12~M$_\mathrm{Jup}$.

In our `cold-start' calculations the track for  Run 2C, 13.7~M$_\mathrm{Jup}$,
passes close to the object in the (log $L$, $t$) diagram, 
and the calculations give $T_\mathrm{eff}=1627$~K at an age of 12~Myr. Our mass
 10~M$_\mathrm{Jup}$ cannot possibly provide a fit.  The `hot-start'
models thus would show that the object is a planet, as defined by an object with
mass not high enough to burn deuterium. However this particular 
 `cold-start' model indicates that beta Pictoris b is presently burning
deuterium, which,
according to the same definition, would classify  it as a brown dwarf.
As mentioned in Section \ref{sect:intro}, this definition is not universally
agreed upon; an alternative definition, based on the minimum in the mass
distribution of low-mass companions, observed within 
several AU of sunlike stars,  places the limit at $\approx 25$~M$_\mathrm{Jup}$. 
In this case Beta Pic b would  still be a planet.
 Note that in the `cold-start' calculations, the fit at 13.7~M$_\mathrm{Jup}$  
with an assumed $\sigma=6\,\mathrm{g\,cm}^{-2}$ is not unique; the
companion could also be fit at $\sigma=4\,\mathrm{g\,cm}^{-2}$ at a slightly 
higher mass, 
about 15.6~M$_\mathrm{Jup}$. Furthermore, these masses are uncertain and
will  probably change  when the calculations are redone in the future with
more detailed model atmospheres. Nevertheless, as such they are marginally
consistent with the upper limits to the mass of Beta Pic b derived from
radial velocity measurements \citep{lag12}. For a planet at 9 AU the limit
is 12~M$_\mathrm{Jup}$; at 10 AU it is 15.4~M$_\mathrm{Jup}$.

We note also that the luminosity curve for 11.7~M$_\mathrm{Jup}$ 
in Figure~\ref{fig:5} agrees
well with the observed luminosity of the directly imaged planet
HR 8799 c at the stellar age ($\approx 6 \times 10^7$ years). The observed
value is given by \citet{mar12} as log $L$/L$_\odot = -4.9  \pm 0.1$.
The agreement of course requires a core mass of $\approx 30$~M$_\oplus$.
A hot-start model of about 10~M$_\mathrm{Jup}$ without a core also agrees.
However we do not make a detailed comparison with HR 8799 c, because
the metallicity of the star is low ([Fe/H] $= -0.47$) and the planet
orbits at 43 AU, making it highly debatable whether it could have formed
by core-nucleated accretion.   

\begin{table*}
 \caption{Summary}\label{table:3}
 \centering
% \vspace*{1ex}%
 %\vskip 0.2 in%
 \resizebox{0.6\textwidth}{!}{%
 \small
 \begin{tabular}{|l||ccccc|c|}
 \hline
 Run
 &  M/M$_\odot$ & Distance (AU) & $\sigma$ ($\mathrm{g\,cm}^{-2}$)  &$M_\mathrm{core}$
(M$_\oplus$)
 & ${M}_{50}$ (M$_\mathrm{Jup}$)
   \\
 \hline\hline
1A 
 &         1   & 5.2 &10 & 16.8  
 &  12.37 
  \\
 \hline
1B           
 &         1   & 5.2 &10  & 16.8
 &  12.20                          
    \\
 \hline
1C
 &         1   & 5.2 &4  & 4.83
 &  13.55                      
   \\
 \hline
2A         
 &         2   & 9.5 &4 & 18.7
 &  11.95                          
    \\
 \hline
2B
 &         2   & 9.5 &4 & 18.8
 &  12.05                         
    \\
 \hline
2C
 &         2   & 9.5 &6 &31.0
 &  11.65                         
   \\
 \hline
 \end{tabular}
       }
\end{table*}
%%%%

\section{Summary and Conclusions}

We investigate the boundary between brown dwarfs and giant planets,
according to the definition that brown dwarfs can burn  the deuterium
that is present when they form,  and giant planets cannot.
The main parameters and the results for $M_{50}$, the boundary
mass at which half of the original deuterium is burned after
4 Gyr, are summarized in Table~\ref{table:3}. The columns give,
respectively, the run identification, the stellar mass in M$_\odot$, $a_p$,
the initial disk solid
surface density $\sigma$  at $a_p$,  the 
resulting $M_\mathrm{core}$,       and $M_{50}$.
The main cases considered involve a planet/brown dwarf at 5.2 AU around a solar-mass star,
and a planet/brown dwarf at 9.5 AU around a star of 2~M$_\odot$.
The table shows that there is only a small variation in the values of
$M_{50}$, which, however, correlate with the core mass in the sense that
the smaller the core mass, the higher the value of $M_{50}$.

The calculations, taken as a whole, indicate that the envelope entropy,
which is a function of initial conditions and which is closely related to the 
core mass through the accretion processes during the formation phase, 
is an important factor in determining $M_{50}$. However, certain physical 
processes during formation are shown to have only a small effect.
Run 1B has the same parameters as Run 1A except that the dust opacity
during the formation phase is higher by a factor that ranges from 2 to 
100, depending on the depth in the envelope.  This
difference has a negligible effect on $M_{50}$. Run 2B has the same
parameters as 2A except that the disk viscosity during the phase of
disk-limited gas accretion is lower by a factor 2.5. This difference
also has a negligible effect on $M_{50}$. However the disk viscosity
is important in another respect. If it is significantly lower than
the range presented here ($\alpha_t \approx 10^{-2}$), then there will
not be time to accrete a planet with mass necessary to burn deuterium
during the lifetime of the disk. The gas accretion rate onto a planet of
4~M$_\mathrm{Jup}$ around a star of 2~M$_\odot$, 
in a disk with $\alpha_t = 4 \times 10^{-4}$, is
reduced by a factor 400 compared with a disk with 
$\alpha_t = 4 \times 10^{-3}$  
\citep{lis09}, corresponding to less than a Jupiter mass in a million years 
for the initial conditions of Run 2B (formation time about 3 Myr). Of
course the minimum viscosity required to build a planet up to about 
12~M$_\mathrm{Jup}$ will depend on parameters such as $\sigma$ and $a_p$.
This question is discussed in more detail in Appendix~\ref{sec:acc-rates-fit}.

\begin{figure}[]
\centering%
\resizebox{\linewidth}{!}{%
\includegraphics[]{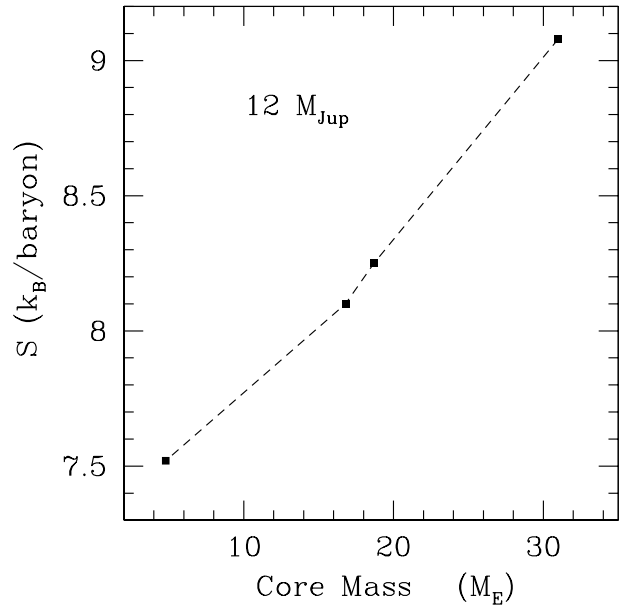}}
\caption{The entropy in the interior of planets of total mass 
12~M$_\mathrm{Jup}$, 
immediately after formation,  is plotted against their core masses, 
in M$_\oplus$. The points plotted, from top to bottom, are from
Runs 2C, 2A, 1B, and 1C.
             }
\label{fig:e}
\end{figure}

\begin{table*}
 \caption{Data for Figure~\ref{fig:e}}\label{table:4}
 \centering
% \vspace*{1ex}%
 %\vskip 0.2 in%
 \resizebox{0.9\textwidth}{!}{%
 \small
 \begin{tabular}{|l||cccccccc|c|}
 \hline
 Run
 &  $M_\mathrm{final}$/M$_\mathrm{Jup}$ & $t_\mathrm{form}$ (yr) & 
$M_\mathrm{core}$/M$_\oplus$ & $T_\mathrm{c,f}$ (K)
 & $T_\mathrm{max}$ (K)  &  $\rho_\mathrm{c,f}$  ($\mathrm{g\,cm}^{-3}$)  & 
 log ($L_\mathrm{f}/\mathrm{L}_\odot$) & $D_\mathrm{final}$/$D_\mathrm{init}$
   \\
 \hline\hline
2C
 &   12.0     & $8.83 \times 10^5$ &31.0 
 &  $3.48 \times 10^5$    & $3.67 \times 10^5$ & 29.6 
 &   -4.80 &$0.278 $
  \\
 \hline
2A 
 &   12.0      & $2.14 \times 10^6$ &18.7 
 &  $2.76 \times 10^5$    & $3.06 \times 10^5$ & 41.8 
 &   -5.63 &$0.435$
  \\
 \hline
1B           
 &  12.0     & $2.66 \times 10^6$ &16.8 
 &  $2.66 \times 10^5$    & $2.66 \times 10^5$ &46.1
 &  -5.91 &$0.930$
    \\
 \hline
1C           
 &  12.0      & $4.10 \times 10^6$ &4.80 
 &  $2.08 \times 10^5$    & $2.08 \times 10^5$ &54.0
 &  -6.56 &$1.000$
    \\
 \hline
 \end{tabular}
       }
\end{table*}
%%%%

Core accretion models, in the cold-start case, are known to have low 
entropy compared with hot-start models. In \citet{mar07} the entropy just
after formation for
10~M$_\mathrm{Jup}$ was found to be 8.2 $k_B$ per baryon for 
$M_\mathrm{core}=  16.8$~M$_\oplus$. The corresponding luminosity at ages
of $10^7$ to $10^8$ years was about $ 2 \times 10^{-6}$ L$_\odot$,
certainly fainter than observed values for directly imaged planets. In 
this mass range, for the  given core mass, 
the entropy is very insensitive to the planet's total
mass, as shown in that paper and confirmed by the present results.
However our calculations show that the entropy is quite sensitive to
the core mass, as illustrated in Figure~\ref{fig:e} (a similar effect
has been found independently by \citet{mor13} for $M_\mathrm{core} 
> 20$~M$_\oplus$).  The points shown
are all calculated with the same total mass and the same  disk viscosity. 
All used the
reduced interstellar grain opacity, except for the point at 
$M_\mathrm{core}= 4.8~$M$_\oplus$, for which the grain-settling 
opacities were used (if the interstellar opacities had been used, the 
formation time would have been considerably longer).
However the comparison between Runs  1A and 1B, which looked at the effect
of changing the opacities, showed that the difference in entropy was
less than 0.1 $k_B$ per baryon at the same total mass. The effect on the entropy
 of changing the viscosity (Runs 2A and 2B) was even smaller. Physical
effects that do affect the entropy include the planetesimal accretion
rate and the rate of contraction of the envelope, both of which
affect the internal heating of the envelope. Thus the
luminosities of newly formed massive planets, depending on formation
conditions, can vary by up to two orders of magnitude.  Information on
the runs whose entropies are plotted in the figure is given in
Table~\ref{table:4}. The table is in the same format as Table~\ref{table:2}
and gives the runs in order of decreasing entropy. Clearly, for this set
of models, a lower entropy is associated with a longer formation
time. The luminosity plots for these four cases in Figure~\ref{fig:lum}
illustrate the same effect.

The combination of $M_\ast$, $a_p$, and $\sigma$
determines the isolation mass,  and thereby the ultimate core
mass, which turns out to be a
key factor in determining the entropy of the planet at formation.
Higher entropy, in particular the higher temperature, 
 favors more rapid nuclear burning, so the higher entropy
runs result in lower values of $M_{50}$.
Nevertheless, the range of initial conditions explored here, which is
considerable,  produces
only a small range in $M_{50}$, about 11.6--13.6~M$_\mathrm{Jup}$, in
agreement with previous independent calculations. We can further conclude,
that for cold-start core-accretion models that do burn deuterium, 
 the tracks in the luminosity
versus time diagram can potentially provide agreement with the properties
of directly-imaged low-mass stellar companions.

\acknowledgments

Primary funding for this project was provided by the NASA Origins of 
Solar Systems Program grant  NNX11AK54G (P. B., G. D., J. L.).
G.D.\ acknowledges additional support  from  NASA 
grant NNX11AD20G. P. B. acknowledges additional support
from NSF grant AST0908807. D. S. is supported in part by NASA grants
NNH11AQ54I and NNH12AT89I.
The authors are indebted to Gilles Chabrier for the use of his nuclear
screening factors.
The 3D hydrodynamical simulations reported in this work were performed
using resources provided by the NASA High-End
Computing (HEC) Program through the NASA Advanced Supercomputing
(NAS) Division at Ames Research Center.
G.D.\ thanks Los Alamos National Laboratory for its hospitality.
The authors thank the referee Dr Christoph Mordasini for  a detailed
and constructive review.

\appendix

\section{Analytic Approximations of the Disk-limited Gas Accretion Rates} \label{sec:acc-rates-fit}

In this section, we provide analytic approximations for the gas 
accretion rate in the regime where this rate is limited by the ability 
of the disk to transfer gas to the planet.
In the calculations, we used piece-wise functions obtained by fitting 
the data from the 3D hydrodynamical calculations 
(see Section~\ref{sec:acc-rates}), for various values 
of the turbulent  viscosity parameter, $\alpha_{t}$, which quantifies
the kinematic viscosity of the disk at the radial location of the planet, 
$\nu_{t}=\alpha_{t} H^{2}_{p} \Omega_{p}$.
We recall that the hydrodynamical calculations used an aspect ratio 
$H_{p}/a_{p}=0.05$, which is a reasonable value in evolved disks 
between $5$ and $10\,\mathrm{AU}$ 
\citep[e.g.,][and references therein]{dm2012}.

We fitted $(\log{\dot{M}_{p}},\log{M_{p}})$ data using 
multiple second-oder polynomials, which were then smoothly joined
in overlapping regions. Since this procedure is somewhat cumbersome,
here we provide simpler analytic approximations derived from data 
in the range of $M_{p}/M_{\star}$ from $10^{-4}$ to $10^{-2}$. 
In the calculations, as explained in Section~\ref{sec:acc-rates}, 
disk-limited accretion sets in when $M_{p}\gtrsim 0.25$~M$_\mathrm{Jup}$.

\begin{figure}[]
\centering%
\resizebox{0.6\linewidth}{!}{%
\includegraphics[]{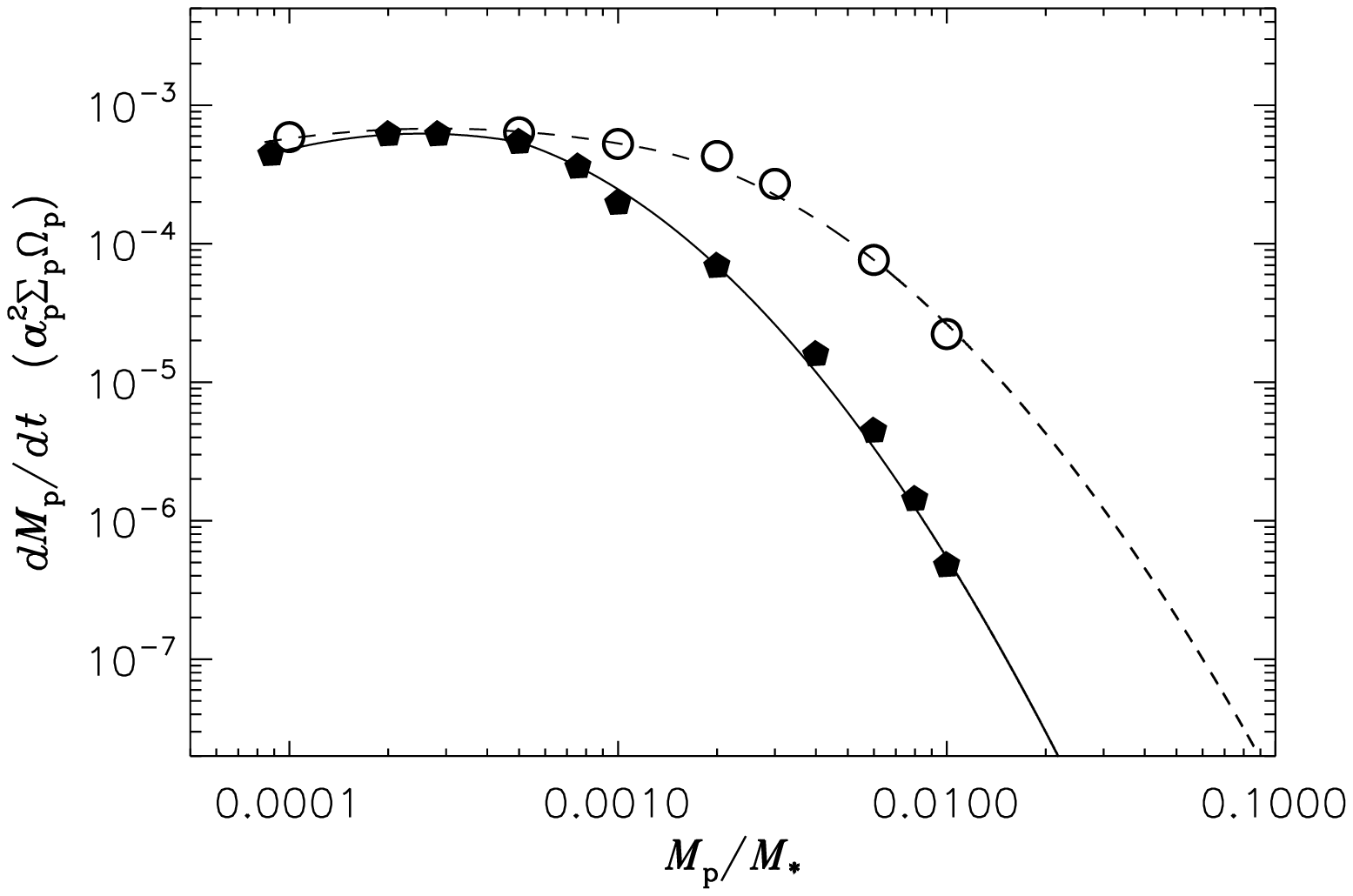}}\\
\resizebox{0.6\linewidth}{!}{%
\includegraphics[]{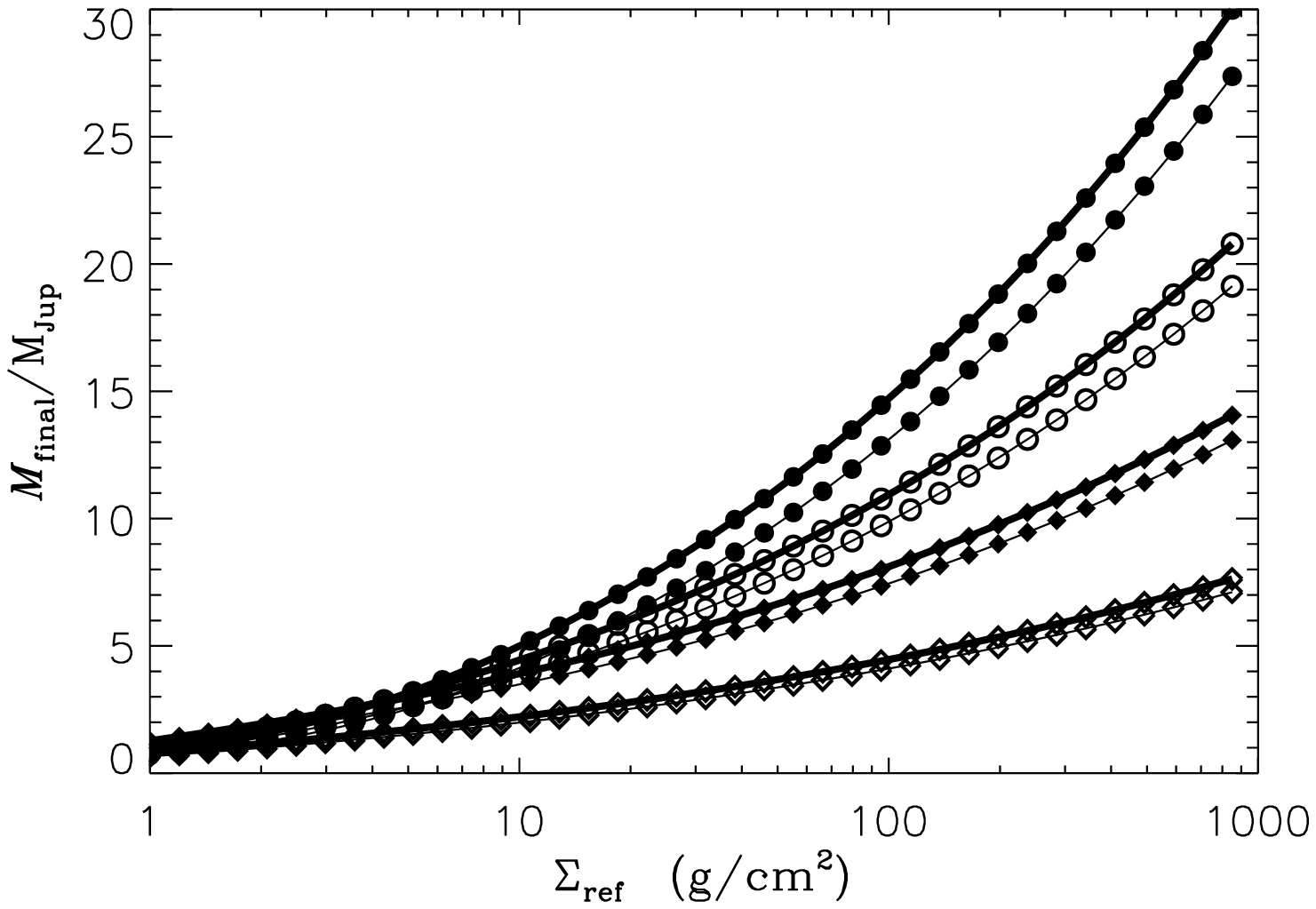}}
\caption{{\it Upper panel}:
              disk-limited gas accretion rates versus the planet-to-star
              mass ratio (see also Figure~\ref{fig:dmdt}). The dashed
              line is Equation~(\ref{eq:fit12}), the solid line is Equation~(\ref{eq:fit43}),
              and the symbols represent the simulations' data for the 
              disk turbulent parameter $\alpha_{t}=10^{-2}$ (open circles)
              and $4\times 10^{-3}$ (filled pentagons).
              {\it Lower panel}:
              final mass of a planet accreting at a disk-limited gas accretion
              rate in various situations: $\alpha_{t}=4\times 10^{-3}$ (diamonds),
              $\alpha_{t}=10^{-2}$ (circles), $a_{p}=5.2$~AU (thin lines),
              $a_{p}=9.5\,\mathrm{AU}$ (thick lines), $M_{\star}=1$~M$_\odot$
              (open  symbols), and $M_{\star}=2$~M$_\odot$ (filled symbols).
              See text for further details.
             }
\label{fig:app}
\end{figure}

Let us introduce the four functions
\begin{eqnarray}
f_{1}(q) & = & a_{0} + a_{1} \log{q} + a_{2}(\log{q})^{2} \label{eq:f1} \\
f_{2}(q) & = & b_{0} + b_{1} \log{q} + b_{2}(\log{q})^{2} \label{eq:f2} \\
f_{3}(q) & = & c_{0} + c_{1} \log{q} + c_{2}(\log{q})^{2} \label{eq:f3} \\
f_{4}(q) & = & d_{0} + d_{1} \log{q} + d_{2}(\log{q})^{2} \label{eq:f4},
\end{eqnarray}
where $q=M_{p}/M_{\star}$ and all logarithms are in base $10$.
The   coefficients $a_{i}, b_{i}, c_{i}$, and $d_{i}$ are given in 
Table~\ref{table:5}.
For $\alpha_{t}=10^{-2}$, the following analytic approximation for 
the disk-limited gas accretion rate, in units of $a^{2}_{p}\Sigma_{p}\Omega_{p}$,
may be used:
\begin{equation}
\log{\dot{M}_{p}}=\left\{%
                    \begin{array}{l l}
                     f_{1}(q) \quad \mathrm{if}\quad q < 0.001197\\
                     f_{2}(q) \quad \mathrm{otherwise}.
                    \end{array}%
                    \right.
\label{eq:fit12}
\end{equation}
For $\alpha_{t}=4\times 10^{-3}$, the following analytic approximation
may be applied
\begin{equation}
\log{\dot{M}_{p}}=\min{[f_{3}(q) ,f_{4}(q)]}.
\label{eq:fit43}
\end{equation}
The fitting functions are shown in the upper panel of Figure~\ref{fig:app}, 
along with the data obtained from the 3D hydrodynamical calculations. 

%%%%
\begin{table}
 \caption{Coefficients in Equations~(\ref{eq:f1})--(\ref{eq:f4})}\label{table:5}
 \centering
% \vspace*{1ex}%
 %\vskip 0.2 in%
 \resizebox{0.5\textwidth}{!}{%
 \small
 \begin{tabular}{|l||ccc|c|}
 \hline

 &  $i=0$ & $i=1$ & $i=2$
   \\
 \hline\hline
$a_{i}$
 & $-6.8345179$ & $-2.5152600$ & $-0.354296$
  \\
 \hline
$b_{i}$ 
 & $-12.471200$ & $-6.3732500$ & $-1.014440$
  \\
 \hline
$c_{i}$       
 & $-11.429400$ & $-5.0171300$ & $-0.697484$
    \\
 \hline
$d_{i}$        
 & $-17.819292$ & $-8.5200885$ & $-1.172181$
    \\
 \hline
 \end{tabular}
       }
\end{table}
%%%%

%

In the range of the turbulent parameter $\alpha_{t}$ that we explored
($4\times 10^{-4}\le \alpha_{t}\le 0.02$),
the maximum of $\dot{M}_{p}$ occurs at a ratio $M_{p}/M_{\star}$ 
similar (within a factor of $\approx 2$) to the square root of the right-hand 
side of Equation~(\ref{eq:tor-con}), i.e., before gas begins to be depleted
significantly because of the formation of the density gap. 
The maximum of $\dot{M}_{p}$ can be compared to the (steady state) 
accretion rate through the disk in absence of the planet, 
$3\pi\nu_{t}\Sigma_{p}$ \citep{lbp1974}, at the radial location of the
planet. 
In units of $a^{2}_{p}\Sigma_{p}\Omega_{p}$, 
this accretion rate can be written as $3\pi\alpha_{t}(H_{p}/a_{p})^2$, 
giving   $\approx 10^{-4}$ and $2.4\times 10^{-4}$ for  $\alpha_{t}$ = 
0.004 and 0.01, respectively.  As can be seen in Figure~\ref{fig:app},
these disk accretion rates are smaller than the maximum of $\dot{M}_{p}$.
However, it should be noted that the tidal perturbation of the planet 
can modify the accretion rate through the disk \citep{ld2006}.

Equations~(\ref{eq:fit12}) and (\ref{eq:fit43}) can be integrated to find
the final (asymptotic) mass of a planet, $M_\mathrm{final}$, that
accretes gas at a disk-limited gas accretion rate. We solved numerically
the differential equation
\begin{equation}
 \dot{M}_{p}=F(M_{p},M_{\star},a_{p},\Sigma_{p},\alpha_{t})
 \label{eq:diff_eq}
\end{equation}
for $M_{p}$, using an adaptive Adams-Bashforth-Moulton method 
of variable order with adaptive step-size and error control
(available through the SLATEC Common Mathematical Library).
Notice from Equation~(\ref{eq:diff_eq}) that, although the dependence
of $\dot{M}_{p}$ on $M_{\star}$, $a_{p}$, and $\Sigma_{p}$ is trivial,
the dependence of $M_{p}(t)$ on those three quantities is not!

During the integration of Equation~(\ref{eq:diff_eq}), we assumed that
$M_{\star}$, $a_{p}$, and $\alpha_{t}$ are constants. In oder to mimic
the viscous evolution of the (unperturbed) gas surface  density at 
the radial position of the planet, $\Sigma_{p}$, we applied the solution 
of \citet{lbp1974} for a disk with no central couple. 
Using the same notations and
indicating with $R_{1}$ the initial standard deviation of the (gaussian)
surface density distribution and with $M_{1}$ the
initial disk mass, \citet{lbp1974} found that
$\nu_{t}/R^{2}_{1}=(2/3)\dot{M}_{*}/M_{1}$ 
(where $\dot{M}_{*}$ is the initial accretion rate on the star).
Introducing the non-dimensional `viscous' time\symbolfootnote[2]{%
Notice that the power of $R_{1}$, in the definition of \citet{lbp1974},
 should be $-2$. Also, the subscript `$1$' in $R_{1}$ and $M_{1}$
 refers to the viscous time $t_{\mathrm{vis}}=1$, when the physical
 time $t=0$.}
$t_{\mathrm{vis}}=6 (\nu_{t}/R^{2}_{1}) t +1$, which can be written as
$t_{\mathrm{vis}}=4 (\dot{M}_{*}/M_{1}) t +1$, 
the surface density evolution can be approximated by 
\begin{equation}
  \Sigma_{p}=\Sigma_{\mathrm{ref}}\, t^{-5/4}_{\mathrm{vis}},
  \label{eq:tvis}
\end{equation}
where $\Sigma_{\mathrm{ref}}$ is a parameter and 
which represents the behavior of the \citeauthor{lbp1974} solution for 
$t_{\mathrm{vis}}\gg 1$.
We assumed that $\dot{M}_{*}/M_{1}$ is 
$\approx 10^{-6}\,\mathrm{yr^{-1}}$.
According to the equation above, the surface density ratio 
$\Sigma_{p}/\Sigma_{\mathrm{ref}}$
decreases by more than two orders of magnitude over $10$~Myr.

We integrated Equation~(\ref{eq:diff_eq}) for the values of 
$M_{\star}$, $a_{p}$, and $\alpha_{t}$ used in the calculations, 
applying Equation~(\ref{eq:tvis}), and determined $M_\mathrm{final}$
as a function of $\Sigma_{\mathrm{ref}}$.
The results are shown in the lower panel of Figure~\ref{fig:app}
(see figure caption for a description of the different curves).
The final mass is reached within about $5.5$~Myr, when typically
$M_{p}/\dot{M}_{p}\sim 100$~Myr. 
The effect of disk viscosity is evident in this figure. 
In fact, around a solar mass star, the mass threshold for deuterium
burning can only be achieved for $\alpha_{t}\gtrsim 10^{-2}$.
Among the varied parameters, $\alpha_{t}$ produces the largest
differences in $M_\mathrm{final}$, whereas $a_{p}$ produces
the smallest.
Notice that the values of $M_\mathrm{final}$ shown in the lower panel
of Figure~\ref{fig:app} should not necessarily agree with those in the 
D-burning calculations because of the different assumptions made for the 
nebula evolution. 
In particular, $\Sigma_{p}$ in those calculations was taken as a constant.

\end{document}